\documentclass[amsmath,amssymb]{revtex4}

\usepackage{graphicx}
\usepackage{amscd,amsmath,amsthm,amsfonts,amssymb}
\def\[{\begin{equation}}
\def\]{\end{equation}}

\begin{document}
\title{Universal rogue wave patterns associated with the Yablonskii-Vorob'ev polynomial hierarchy}
\author{Bo Yang and Jianke Yang}
\affiliation{Department of Mathematics and Statistics, University of Vermont, Burlington, VT 05405, USA}
\begin{abstract}
We show that universal rogue wave patterns exist in integrable systems. These rogue patterns comprise fundamental rogue waves arranged in shapes such as triangle, pentagon and heptagon, with a possible lower-order rogue wave at the center. These patterns appear when one of the internal parameters in bilinear expressions of rogue waves gets large. Analytically, these patterns are determined by the root structures of the Yablonskii-Vorob'ev polynomial hierarchy through a linear transformation. Thus, the induced rogue patterns in the space-time plane are simply the root structures of Yablonskii-Vorob'ev polynomials under actions such as dilation, rotation, stretch, shear and translation. As examples, these universal rogue patterns are explicitly determined and graphically illustrated for the generalized derivative nonlinear Schr\"odinger equations, the Boussinesq equation, and the Manakov system. Similarities and differences between these rogue patterns and those reported earlier in the nonlinear Schr\"odinger equation are discussed.
\end{abstract}

\maketitle

\section{Introduction}
Rogue waves were first studied in oceanography, where they referred to large spontaneous and localized water wave excitations that are a threat even to big ships \cite{Ocean_rogue_review,Pelinovsky_book}. Later, their counterparts in optics were also reported \cite{Solli_Nature,Wabnitz_book}. Due to their physical importance, rogue waves have received intensive theoretical and experimental studies in the past decade. One physical mechanism for rogue wave generation is modulation instability of nonlinear periodic wave trains --- also called Benjamin-Feir instability in water waves \cite{Benjamin}. Under this mechanism, the first rogue wave solution was derived by Peregrine for the nonlinear Schr\"{o}dinger (NLS) equation \cite{Peregrine}, which governs wave envelope propagation in the ocean and optical systems \cite{Benney}. In this NLS equation for wave envelopes, the modulation instability of periodic wave trains reduces to the instability of uniform (envelope) background. Peregrine's solution was later generalized to higher orders which could reach higher peak amplitudes \cite{AAS2009,DGKM2010,KAAN2011,GLML2012,OhtaJY2012}. Since modulation instability also arises in many other integrable physical systems, rogue waves have also been derived for such systems, such as the derivative NLS equations for circularly polarized nonlinear Alfv\'en waves in plasmas and short-pulse propagation in a frequency-doubling crystal
\cite{Kaup_Newell,KN_Alfven1,KN_Alfven2,Wise2007,KN_rogue_2011,KN_rogue_2013,CCL_rogue_Chow_Grimshaw2014,CCL_rogue_2017,YangDNLS2019}, the Manakov equations for light transmission in randomly birefringent fibers \cite{Menyuk,BDCW2012,ManakovDark,LingGuoZhaoCNLS2014,Chen_Shihua2015,ZhaoGuoLingCNLS2016}, and the three-wave resonant interaction equations \cite{Ablowitz_book,BaroDegas2013,DegasLomba2013,ChenSCrespo2015,WangXChenY2015,ZhangYanWen2018}. Experimentally, various rogue waves governed by the NLS equation and defocusing Manakov equations have been observed in optical fibers, water tanks, and plasma \cite{Fiber1,Fiber2,Fiber3,Tank1,Tank2,Plasma}.

Pattern formation in rogue waves is an important question, as such information allows for the prediction of later rogue wave events from earlier wave forms. For the NLS equation, this question has been investigated in \cite{KAAN2011,HeFokas,KAAN2013,Akhmediev_triangular,YangNLS2021}. It was observed in \cite{KAAN2011} that if a $N$-th order rogue wave exhibits a single-shell ring structure, then the center of the ring is a $(N-2)$-th order rogue wave. This observation was explained analytically in \cite{HeFokas} through Darboux transformation. In \cite{KAAN2013}, it was observed that NLS rogue patterns could be classified according to the order of the rogue waves and the parameter shifts applied to the Akhmediev breathers in the rogue-wave limit. This latter observation allowed the authors to extrapolate the shapes of rogue waves beyond order six, where numerical plotting of rogue waves became difficult. In \cite{Akhmediev_triangular}, it was stated that the total number of elementary (Peregrine) rogue waves in a $N$-th order rogue wave solution was $N(N + 1)/2$ (but those elementary rogue waves were allowed to coalesce).
In \cite{YangNLS2021}, we predicted analytically that when one of the internal parameters in a high-order NLS rogue wave was large, then this rogue wave would exhibit clear geometric patterns, comprising Peregrine waves forming shapes such as triangle, pentagon and heptagon, with a possible lower-order rogue wave at its center. Asymptotically, this rogue pattern was given by the root structure of the Yablonskii-Vorob'ev polynomial hierarchy through dilation and rotation. This deep connection between rogue patterns and root structures of Yablonskii-Vorob'ev polynomials is a drastic step forward in our understanding of rogue events. As small applications of our analytical theory, all previous numerical observations and heuristic statements in \cite{KAAN2011,KAAN2013,Akhmediev_triangular} were explained.

A natural question that follows is, do these NLS rogue patterns associated with the Yablonskii-Vorob'ev polynomial hierarchy also appear in other integrable systems? The answer is a definite yes. In this paper, we study rogue patterns in three other integrable systems --- namely, the generalized derivative nonlinear Schr\"odinger (GDNLS) equations, the Boussinesq equation, and the Manakov system. In all these three systems, we show analytically and verify numerically that when one of the internal parameters in their rogue wave solutions is large, then these rogue waves would also exhibit similar geometric patterns that are described by the root structures of the Yablonskii-Vorob'ev hierarchy, with a possible lower-order rogue wave at the center. The main difference from the NLS rogue patterns is that, while the NLS rogue pattern is just a dilation and rotation of the root structure of Yablonskii-Vorob'ev polynomials, rogue patterns in these three other integrable systems may also involve other actions such as stretch, shear and translation to the Yablonskii-Vorob'ev root structure. All these actions from the Yablonskii-Vorob'ev root structure to the rogue pattern are analytically described by a linear transformation, the details of which depend on the underlying integrable equation. We thus conclude that, there exist in integrable systems universal rogue patterns, which are the Yablonskii-Vorob'ev root structures under linear transformations.

This paper is structured as follows. In Sec.~\ref{sec:pre}, bilinear expressions of rogue waves are presented for the GDNLS, Bousinesq and Manakov equations. In addition, the Yablonskii-Vorob'ev polynomial hierarchy is introduced, and its root structures displayed. In Sec.~\ref{s:prediction}, we present our analytical predictions of rogue patterns in these three integrable systems when one of their internal parameters is large. All these rogue patterns turn out to be just the Yablonskii-Vorob'ev root structures under linear transformations. In Sec.~\ref{sec:comparison}, these analytical predictions of rogue patterns are compared to true rogue wave solutions, and complete agreement is demonstrated. Sec.~\ref{sec:proof} contains proofs for the analytical predictions of rogue patterns presented in Sec.~\ref{s:prediction}. In Sec.~\ref{sec:universal}, we describe the common features of rogue patterns in these three integrable systems and others, and highlight the universality of such patterns associated with the Yablonskii-Vorob'ev hierarchy. Sec.~\ref{sec:summary} summarizes the results of this paper. The two appendices contain brief derivations of bilinear rogue waves given in Sec.~\ref{sec:pre} for the Boussinesq and Manakov equations.

\section{Preliminaries} \label{sec:pre}
In this article, we will consider three integrable systems and show that their rogue waves exhibit universal patterns similar to those reported in Ref.~\cite{YangNLS2021} for the NLS equation. For that purpose, we will present these three integrable systems as well as their bilinear rogue wave solutions first. These three integrable systems are the GDNLS equations, the Boussinesq equation, and the Manakov system. We note that although we have derived bilinear rogue waves for the GDNLS equations and the Boussinesq equation before \cite{YangDNLS2019,BoussiRWs2020}, those bilinear expressions can be further simplified, and it is these simplified bilinear expressions which we will present below. For the Manakov system, however, bilinear expressions of rogue waves have not been derived before; so our solutions to be presented in this section will be new results. Although rogue waves for the Manakov system have been derived by Darboux transformation in Refs.~\cite{BDCW2012,ManakovDark,LingGuoZhaoCNLS2014,Chen_Shihua2015,ZhaoGuoLingCNLS2016}, those rogue expressions from Darboux transformation are not convenient for our rogue-pattern analysis and thus will not be used.

Our bilinear rogue waves will be presented through elementary Schur polynomials. These Schur polynomials $S_j(\mbox{\boldmath $x$})$, with $\emph{\textbf{x}}=\left( x_{1}, x_{2}, \ldots \right)$, are defined by
\begin{equation} \label{Schurdef}
\sum_{j=0}^{\infty}S_j(\mbox{\boldmath $x$}) \epsilon^j
=\exp\left(\sum_{k=1}^{\infty}x_k \epsilon^k\right),
\end{equation}
or more explicitly,
\begin{equation} \label{Skdef}
S_{j}(\mbox{\boldmath $x$}) =\sum_{l_{1}+2l_{2}+\cdots+ml_{m}=j} \left( \ \prod _{k=1}^{m} \frac{x_{k}^{l_{k}}}{l_{k}!}\right).
\end{equation}

\subsection{The GDNLS equations and their rogue wave solutions}
The normalized GDNLS equations are \cite{Kundu1984,Clarkson1987,Satsuma_GDNLS_soliton,YangDNLS2019}
\[ \label{GDNLS}
\textrm{i}u_t+\frac{1}{2}u_{xx}+\textrm{i}\gamma |u|^2u_x+\textrm{i}(\gamma-1)u^2u_x^*+\frac{1}{2}(\gamma-1)(\gamma-2)|u|^4u=0,
\]
where $\gamma$ is a real constant. These equations become the Kaup-Newell equation when $\gamma=2$ \cite{Kaup_Newell}, the Chen-Lee-Liu equation when $\gamma=1$ \cite{CCL}, and the Gerdjikov-Ivanov equation when $\gamma=0$ \cite{GI}. These GDNLS equations and their special versions govern a number of physical processes such as the propagation of circularly polarized nonlinear Alfv\'en waves in plasmas \cite{KN_Alfven1,KN_Alfven2}, short-pulse propagation in a frequency-doubling crystal \cite{Wise2007}, and propagation of ultrashort pulses in a single-mode optical fiber \cite{Agrawal_book,Kivshar_book}. Rogue waves in these equations satisfy the following normalized boundary conditions \cite{YangDNLS2019}
\begin{eqnarray}
&& u(x,t)\rightarrow  e^{{\rm i}(1-\gamma-\alpha) x-{\frac{\rm i}{2}}\left[\alpha^2+2(\gamma-2)\alpha+1-\gamma\right]t},  \quad x, t \to \pm \infty, \label{BoundaryCond1}
\end{eqnarray}
where $\alpha>0$ is a free wave number parameter. Unlike the NLS equation, the GDNLS equations (\ref{GDNLS}) do not admit Galilean invariance. Thus, $\alpha$ is an irreducible parameter in these rogue waves.

General rogue waves to the GDNLS equations (\ref{GDNLS}) have been derived by the bilinear method in Ref.~\cite{YangDNLS2019}. It turns out that those rogue expressions can be further simplified by setting all $(x_{2}^{\pm}, x_{4}^{\pm}, x_{4}^{\pm}, \cdots)$ to zero, similar to the NLS case we reported in Ref.~\cite{YangNLS2021} for similar reasons. In addition, we can set the first parameter $a_1$ to zero by a shift of $(x,t)$. Then, we get the following lemma for bilinear rogue waves in the GDNLS equations.

\begin{quote}
\textbf{Lemma 1.} \emph{The GDNLS equations (\ref{GDNLS}) under the boundary condition  (\ref{BoundaryCond1}) admit rogue wave solutions}
\begin{eqnarray}
&& u_N(x,t)= e^{{\rm i}(1-\gamma-\alpha) x
-{\frac{\rm i}{2}}\left[\alpha^2+2(\gamma-2)\alpha+1-\gamma\right]t}\hspace{0.05cm}\frac{(f_N^*)^{\gamma-1}g_N}{f_N^\gamma}, \label{BilinearTrans2}
\end{eqnarray}
\emph{where the positive integer $N$ represents the order of the rogue wave, `*' represents complex conjugation,}
\begin{equation}
f_N(x,t)=\sigma_{0,0}, \quad g_N(x,t)=\sigma_{-1,1},
\end{equation}
\begin{equation} \label{sigma_nD}
\sigma_{n,k}=
\det_{\begin{subarray}{l}
1\leq i, j \leq N
\end{subarray}}
\left(\begin{array}{c}
\phi_{2i-1,2j-1}^{(n,k)}
\end{array}\right),
\end{equation}
\emph{the matrix elements in $\sigma_{n,k}$ are defined by}
\[ \label{matrixmnij}
\phi_{i,j}^{(n,k)}=\sum_{\nu=0}^{\min(i,j)} \frac{1}{4^{\nu}} \hspace{0.06cm} S_{i-\nu}(\textbf{\emph{x}}^{+}(n,k) +\nu \textbf{\emph{s}})  \hspace{0.06cm} S_{j-\nu}(\textbf{\emph{x}}^{-}(n,k) + \nu \textbf{\emph{s}}),
\]
\emph{vectors} $\textbf{\emph{x}}^{\pm}(n,k)=\left( x_{1}^{\pm}, x_{2}^{\pm},\cdots \right)$ \emph{are defined by}
\begin{eqnarray}
&&x_{1}^{+}=\hspace{0.2cm}   k+ \left(n+\frac{1}{2}\right)\left(h_{1}+\frac{1}{2}\right)+\sqrt{\alpha}x+ \sqrt{\alpha}\left[(\alpha-1)+ {\rm i}\sqrt{\alpha}\right]t,  \\
&&x_{1}^{-}=-k- \left(n+\frac{1}{2}\right)\left(h_{1}^*+\frac{1}{2}\right)+\sqrt{\alpha}x+ \sqrt{\alpha}\left[(\alpha-1)-{\rm i}\sqrt{\alpha}\right]t, \\
&&x_{2r+1}^{+}=\hspace{0.25cm} (n+\frac{1}{2}) \hspace{0.04cm} h_{2r+1}+ \frac{1}{(2r+1)!}\left\{\sqrt{\alpha}x+\left[\sqrt{\alpha}(\alpha-1)+2^{2r} {\rm i} \alpha\right]t\right\} +a_{2r+1},  \quad r\ge 1,  \\
&&x_{2r+1}^{-}= -(n+\frac{1}{2}) \hspace{0.04cm} h_{2r+1}^*+ \frac{1}{(2r+1)!}\left\{\sqrt{\alpha}x+\left[ \sqrt{\alpha}(\alpha-1)-2^{2r} {\rm i} \alpha\right]t\right\} + a_{2r+1}^*, \quad r\ge 1, \\
&& x_{2r}^{\pm}=0, \quad r\ge 1,
\end{eqnarray}
\emph{$\emph{\textbf{s}}=(s_{1}, s_{2}, \cdots)$, $h_{r}(\alpha)$ and $s_{r}$ are coefficients from the expansions}
\begin{eqnarray}
&& \sum_{r=1}^{\infty} h_{r}\lambda^{r}=\ln \left(\frac{{\rm i}e^{\lambda/2}+\sqrt{\alpha} e^{-\lambda/2}}{{\rm i}+\sqrt{\alpha}}\right),  \quad \sum_{r=1}^{\infty} s_{r}\lambda^{r}=\ln \left[\frac{2}{\lambda}  \tanh \left(\frac{\lambda}{2}\right)\right],
\end{eqnarray}
\emph{and $a_{3}, a_{5}, \cdots, a_{2N-1}$ are free irreducible complex constants.}
\end{quote}

\subsection{The Boussinesq equation and its rogue wave solutions}
The Boussinesq equation was introduced in 1871 for the propagation of long surface waves on water of constant depth \cite{JBoussinesq1871,JBoussinesq1872} (see also \cite{FUrsell1853}). After variable normalizations and variable shifts, this equation can be written as \cite{Clarkson_Boussi,BoussiRWs2020}
\[ \label{BoussiEq}
u_{tt}+u_{xx}-(u^2)_{xx}-\frac{1}{3}u_{xxxx}=0,
\]
and its rogue waves are subject to the boundary conditions
\[ \label{BoundCondition}
u(x,t) \to 0, \quad x, t \to \pm \infty.
\]
Special rogue waves in this equation were considered in \cite{Clarkson_Boussi,Tajiri1,Rao2017}. General rogue waves were derived in \cite{BoussiRWs2020}; however, bilinear expressions of those general rogue waves could be significantly simplified through a new parameterization \cite{YangDNLS2019}. The simplified bilinear rogue waves are presented in the following lemma, and their derivation will be provided in Appendix~A.

\begin{quote}
\textbf{Lemma 2.}
\emph{General rogue waves in the Boussinesq equation (\ref{BoussiEq}) under the boundary condition (\ref{BoundCondition}) are }
\[  \label{Boussi_rogue}
u_{N}(x,t)=2 \partial_{x}^2 \ln \sigma ,
\]
\emph{where}
\[ \label{SigmanAlg}
\sigma(x,t)=\det_{
\begin{subarray}{l}
1\leq i, j \leq N
\end{subarray}
}
\left(  \phi_{2i-1,2j-1} \right),
\]
\[ \label{matrixmij}
\phi_{i,j}=\sum_{\nu=0}^{\min(i,j)} \left(\frac{-1}{12}\right)^{\nu} \hspace{0.06cm} S_{i-\nu}(\textbf{\emph{x}}^{+} +\nu \textbf{\emph{s}})  \hspace{0.06cm} S_{j-\nu}(\textbf{\emph{x}}^{-} + \nu \textbf{\emph{s}}),
\]
\emph{vectors} $\textbf{\emph{x}}^{\pm}=\left(  x_{1}^{\pm}, x_{2}^{\pm},\cdots \right)$ \emph{and} $\emph{\textbf{s}}=(s_{1}, s_{2}, \cdots)$ \emph{are defined by}
\begin{eqnarray}
&&x_{2r+1}^{+}=\frac{\sqrt{3}\hspace{0.04cm}\textrm{i}}{2\cdot 3^{2r+1}\cdot (2r+1)!} \hspace{0.03cm} \left( x + 2^{2r} \textrm{i}t \right)+a_{2r+1}, \quad r=0, 1, 2, \cdots, \label{skrkexpcoeffBoussi1}\\
&&x_{2r+1}^{-}=\frac{\sqrt{3}\hspace{0.04cm}\textrm{i}}{2\cdot 3^{2r+1}\cdot (2r+1)!} \hspace{0.03cm} \left( x - 2^{2r} \textrm{i}t \right) -a_{2r+1}^*, \quad r=0, 1, 2, \cdots, \label{skrkexpcoeffBoussi2}\\
&&x_{2r}^{\pm}=0, \quad r=1, 2, 3, \cdots, \\
&& \sum_{r=1}^{\infty} s_{r}\lambda^{r}=\ln\left[ \frac{2\textrm{i}\sqrt{3}}{\lambda} \tanh\frac{\lambda}{6}\tanh\left(\frac{\lambda}{6}+ \frac{2\textrm{i}\pi}{3}\right) \right], \label{skrkexpcoeffBoussi}
\end{eqnarray}
\emph{$a_{1}=0$, and $a_{3}, a_{5}, \cdots, a_{2N-1}$ are free irreducible complex constants.}
\end{quote}

\textbf{Note 1.} In these Boussinesq rogue waves, we have normalized $a_1=0$ through a shift of $(x,t)$ axes. But we still keep $a_1$ in the solution formulae for reasons which we will explain in Theorem 2 of the next section. A similar treatment will be applied to the Manakov system as well, see Lemma 3 below.

\textbf{Note 2.} The function $f(\lambda)$ on the right side of Eq.~(\ref{skrkexpcoeffBoussi}) satisfies the symmetry $f^*(\lambda)=f(-\lambda)$, where $\lambda$ is considered real. Because of that, all $s_{even}$ values are real, and all $s_{odd}$ values are purely imaginary. The first few $s_k$ values are
\begin{equation} \label{skB}
s_1=\frac{2 \rm{i}}{3 \sqrt{3}}, \quad s_2=-\frac{5}{108}, \quad s_3=-\frac{5 \rm{i}}{243 \sqrt{3}}.
\end{equation}

\subsection{The Manakov system and its rogue wave solutions}

The Manakov system is \cite{Manakov}
\begin{eqnarray}
&&  \left(\textrm{i} \partial_t+ \partial_x^2 \right)u_1+\left(\epsilon_{1}|u_{1}|^2+\epsilon_{2}|u_{2}|^2\right)u_{1}=0, \label{ManakModel1} \\
&& \left(\textrm{i} \partial_t+ \partial_x^2 \right)u_{2}+\left(\epsilon_{1}|u_{1}|^2+\epsilon_{2}|u_{2}|^2\right)u_{2}=0, \label{ManakModel2}
\end{eqnarray}
where $\epsilon_{1}=\pm 1$ and $\epsilon_{2}=\pm 1$. These equations govern many physical processes such as the interaction of two incoherent light beams in crystals \cite{Kivshar_book,Stegeman_Manakov,Segev_coupled}, the transmission of light in a randomly birefringent optical fiber \cite{ManakovPMD1,ManakovPMD2,Wabnitzexperiment1,Wabnitzexperiment2}, and the evolution of two-component Bose-Einstein condensates \cite{BEC,BEC_Manakov_experiment}. This system admits plane wave solutions
\begin{eqnarray}
&& u_{1,0}(x,t)= \rho_{1}  e^{{\rm{i}} (k_{1}x + \omega_{1} t)},    \\
&& u_{2,0}(x,t)= \rho_{2}  e^{{\rm{i}} (k_{2}x + \omega_{2} t)},
\end{eqnarray}
where $(k_1, k_2)$ and $(\omega_1, \omega_2)$ are the wavenumbers and frequencies of the two components, and $(\rho_1, \rho_2)$ are their amplitudes which will be set real using phase invariance of the system. Parameters of these plane waves satisfy the following relations,
\begin{eqnarray}
&& \omega_{1} = \epsilon_{1}\rho_{1}^2 +  \epsilon_{2}\rho_{2}^2 -k_{1}^2,  \\
&& \omega_{2} = \epsilon_{1}\rho_{1}^2 +  \epsilon_{2}\rho_{2}^2 -k_{2}^2.
\end{eqnarray}
Then, boundary conditions for rogue waves in the Manakov system are
\begin{eqnarray}
 u_{j}(x,t)\rightarrow u_{j,0}(x,t),  \ \ \ x, t\to \pm \infty, \quad j=1, 2. \label{BoundaryCond}
\end{eqnarray}

Rogue waves in the Manakov system have been derived in Refs.~\cite{BDCW2012,ManakovDark,LingGuoZhaoCNLS2014,Chen_Shihua2015,ZhaoGuoLingCNLS2016}, all by Darboux transformation. However, those rogue expressions are inconvenient for our rogue pattern analysis. Thus, we have derived bilinear rogue expressions, which will be presented in the following lemma. Details of this derivation will be provided in Appendix B.

\begin{quote}
\textbf{Lemma 3.} \hspace{0.05cm} \emph{If the algebraic equation $\mathcal{F}_{1}'(p)=0$, where}
\[ \label{ManakQ1poly}
\mathcal{F}_{1}(p)=\frac{\epsilon_{1}\rho_{1}^2}{p-\textrm{i} k_{1}}+\frac{\epsilon_{2}\rho_{2}^2}{p-\textrm{i} k_{2}}+2p
\]
\emph{and the prime denotes differentiation, admits a non-imaginary simple root $p_0$, then the Manakov system (\ref{ManakModel1})-(\ref{ManakModel2}) under boundary conditions (\ref{BoundaryCond}) admits rogue wave solutions}
\begin{equation}
u_{1,N}(x,t)= \rho_{1}\frac{g_{1,N}}{f_{N}} e^{{\rm{i}} (k_1x+\omega_{1} t)}, \quad  u_{2,N}(x,t)= \rho_{2}\frac{g_{2,N}}{f_{N}} e^{{\rm{i}} (k_{2}x + \omega_{2} t)}, \label{u1u2}
\end{equation}
\emph{where $N$ is an arbitrary positive integer which represents the order of the rogue wave,}
\[ \label{SchpolysolufN}
f_{N}=\sigma_{0,0}, \quad g_{1,N}=\sigma_{1,0}, \quad  g_{2,N}=\sigma_{0,1},
\]
\[ \label{sigmank1}
\sigma_{n,k}=
\det_{
\begin{subarray}{l}
1\leq i, j \leq N
\end{subarray}
}
\left(
\begin{array}{c}
\phi_{2i-1,2j-1}^{(n,k)}
\end{array}
\right),
\]
\emph{the matrix elements in $\sigma_{n,k}$ are defined by}
\[ \label{Schmatrimnij}
\phi_{i,j}^{(n,k)}=\sum_{\nu=0}^{\min(i,j)} \left[ \frac{|p_{1}|^2 }{(p_{0}+p_{0}^*)^2}  \right]^{\nu} \hspace{0.06cm} S_{i-\nu}(\textbf{\emph{x}}^{+}(n,k) +\nu \textbf{\emph{s}})  \hspace{0.06cm} S_{j-\nu}(\textbf{\emph{x}}^{-}(n,k) + \nu \textbf{\emph{s}}^*),
\]
\emph{vectors} $\textbf{\emph{x}}^{\pm}(n,k)=( x_{1}^{\pm}, x_{2}^{\pm},\cdots )$ \emph{are defined by}
\begin{eqnarray}
&&x_{r}^{+}(n,k)= p_r x +  \left(\sum _{i=0}^r p_i p_{r-i}\right)  (\textrm{i} t) +  n \theta_{r} + k \lambda_{r}   +a_{r},  \quad \mbox{when $r$ is odd,}  \label{defxrp} \\
&&x_{r}^{-}(n,k)= p^*_r x - \left( \sum _{i=0}^r p^*_i p^*_{r-i} \right) (\textrm{i} t)  - n \theta_{r}^* -  k \lambda_{r}^* +a_{r}^*, \hspace{0.25cm} \mbox{when $r$ is odd,} \label{defxrm} \\
&& x_{r}^\pm (n,k)=0, \hspace{5.55cm} \mbox{when $r$ is even,}
\end{eqnarray}
\emph{$\textbf{\emph{s}}=(s_1, s_2, \cdots)$, ($\theta_{r}$, $\lambda_{r}$, $s_{r}$) are coefficients from the expansions}
\begin{eqnarray}
&&  \ln \left[\frac{ p \left( \kappa \right)-\textrm{i} k_{1}}{p_{0}-\textrm{i} k_{1}}\right]  =\sum_{r=1}^{\infty} \theta_{r}\kappa^{r},  \quad
  \ln \left[\frac{ p \left( \kappa \right)-\textrm{i} k_{2}}{p_{0}-\textrm{i} k_{2}}\right]  =\sum_{r=1}^{\infty} \lambda_{r}\kappa^{r}, \label{schucoeflambda} \\
&& \ln \left[\frac{1}{\kappa} \left(\frac{p_{0}+p_{0}^*}{p_{1}} \right) \left( \frac{ p \left( \kappa \right)-p_{0}}{p \left( \kappa \right)+p_{0}^*} \right)  \right] = \sum_{r=1}^{\infty}s_{r} \kappa^r,  \label{schurcoeffsr}
\end{eqnarray}
\emph{the function $p \left(\kappa\right)$ is defined by the equation}
\[\label{defpk}
\mathcal{F}_{1}\left[p \left( \kappa \right)\right] = \mathcal{F}_{1}(p_{0}) \cosh(\kappa),
\]
$p_1\equiv p'(0)$, \emph{$a_{1}=0$, and $a_{3}, a_{5}, \cdots, a_{2N-1}$ are free irreducible complex constants.}
\end{quote}

\textbf{Note 3.} The rogue waves given in Lemma 3 for a simple root in the $\mathcal{F}_{1}'(p)=0$ equation are an important class of rogue waves in the Manakov system, and their $\tau$ functions (\ref{sigmank1}) involve Schur polynomials with index jumps of 2. But such rogue waves are not the only ones in the system. For example,
the Manakov system also admits rogue waves that correspond to a double root in the $\mathcal{F}_{1}'(p)=0$ equation, similar to the three-wave resonant interaction system as studied in Ref.~\cite{YangYang2020-3waves}. In this article, we only consider Manakov rogue patterns in solutions of Lemma 3 associated with a simple root in the $\mathcal{F}_{1}'(p)=0$ equation.

\subsection{The Yablonskii-Vorob'ev polynomial hierarchy} \label{sec:YV}
As for the NLS equation \cite{YangNLS2021}, we will show that rogue patterns in the above three integrable systems are also linked to the root structures of the Yablonskii-Vorob'ev polynomial hierarchy. This polynomial hierarchy has been presented in \cite{Kajiwara-Ohta1996,Clarkson2003-II,YangNLS2021} before. Due to its importance to our work, we present it again below. In addition, a useful formula on its polynomial structure as derived in \cite{YangNLS2021} is also reproduced here.

Let $p_{j}^{[m]}(z)$ be the polynomial defined by
\[ \label{pkmz}
\sum_{j=0}^{\infty}p_j^{[m]}(z) \epsilon^j =\exp\left( z \epsilon - \frac{2^{2m}}{2m+1}\epsilon^{2m+1} \right),\
\]
where $m=1,2,3,\cdots$ is a positive integer. Notice that
\[  \label{pkpkp1}
p_j^{[m]}(z)=[p_{j+1}^{[m]}]'(z).
\]
Then the Yablonskii-Vorob'ev polynomial hierarchy $Q_{N}^{[m]}(z)$ is defined through the $N \times N$ determinant \cite{Kajiwara-Ohta1996,Clarkson2003-II}
\begin{eqnarray} \label{GeneralYablonski}
&& Q_{N}^{[m]}(z) = c_{N} \left| \begin{array}{cccc}
         p^{[m]}_{1}(z) & p^{[m]}_{0}(z) & \cdots &  p^{[m]}_{2-N}(z) \\
         p^{[m]}_{3}(z) & p^{[m]}_{2}(z) & \cdots &  p^{[m]}_{4-N}(z) \\
        \vdots& \vdots & \vdots & \vdots \\
         p^{[m]}_{2N-1}(z) & p^{[m]}_{2N-2}(z) & \cdots &  p^{[m]}_{N}(z)
       \end{array}
 \right|,
\end{eqnarray}
where $p_j^{[m]}\equiv 0$ when $j<0$, and $c_{N}=\prod_{j=1}^{N}(2j-1)!!$. This is a Wronskian determinant in view of the relation (\ref{pkpkp1}). When $m=1$, $Q_{N}^{[1]}(z)$ is the original Yablonskii-Vorob'ev polynomial \cite{Yablonskii1959,Vorobev1965}; and when $m>1$, $Q_{N}^{[m]}(z)$ gives the higher members of the Yablonskii-Vorob'ev hierarchy.

These Yablonskii-Vorob'ev polynomials have the general functional form \cite{YangNLS2021}
\[ \label{QNmform}
Q_{N}^{[m]}(z)=z^{N_0(N_0+1)/2}q_{N}^{[m]}(\zeta), \quad \zeta \equiv z^{2m+1},
\]
where $q_{N}^{[m]}(\zeta)$ is a monic polynomial with all-real coefficients and a nonzero constant term, and the integer $N_0$ is determined uniquely from $(N, m)$ by one of the following two equations
\begin{eqnarray}
&&N\equiv N_{0} \hspace{0.1cm} \mbox{mod} \hspace{0.1cm} (2m+1),    \label{N01}\\
&&N\equiv -N_{0}-1 \hspace{0.1cm} \mbox{mod} \hspace{0.1cm} (2m+1),   \label{N02}
\end{eqnarray}
under the restriction of $0\leq  N_{0} \leq m$. This $N_0$ value is unique since these two equations under that $N_0$ restriction are either mutually-exclusive or yielding the same $N_0$ value. For example, when $N=5$ and $m=4$, the $N_0$ value under the restriction of $0\le N_0\le 4$ can only be found from Eq. (\ref{N02}) as $N_0=3$. Eq.~(\ref{QNmform}) gives the multiplicity of the root zero in $Q_{N}^{[m]}(z)$ as $N_0(N_0+1)/2$ (if $N_0=0$, then zero is not a root). It also shows that the root structure of $Q_{N}^{[m]}(z)$ is invariant under a $2\pi/(2m+1)$-angle rotation in the
complex $z$ plane.

Regarding nonzero roots of $Q_{N}^{[m]}(z)$, it was shown in \cite{Fukutani} that for the original Yablonskii-Vorob'ev polynomials $Q_{N}^{[1]}(z)$, all roots are simple. For the higher Yablonskii-Vorob'ev polynomial hierarchy $Q_{N}^{[m]}(z)$, it was conjectured in \cite{Clarkson2003-II} that all nonzero roots are simple. We have checked this conjecture for a myriad of $(N, m)$ values and found it to hold in all our cases. Thus, we will assume it true in this article. In view of Eq.~(\ref{QNmform}), this implies that the polynomial $Q_{N}^{[m]}(z)$ has
\[ \label{Np}
N_p=\frac{1}{2}\left[N(N+1)-N_{0}(N_{0}+1)\right]
\]
nonzero simple roots.

\begin{figure}[htb]
\begin{center}
\vspace{-5.0cm}
\hspace{-1cm}
\includegraphics[scale=0.9, bb=0 0 385 567]{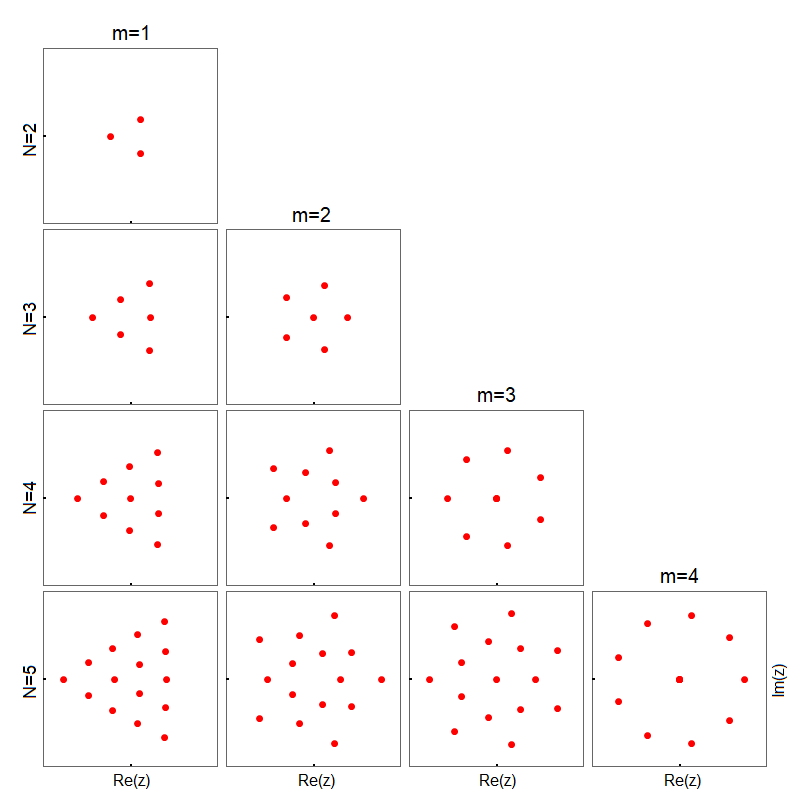}
\caption{Plots of the roots of the Yablonskii-Vorob'ev polynomial hierarchy $Q_{N}^{[m]}(z)$ in the complex $z$ plane for $2\le N\le 5$ and $1\le m\le N-1$. In all panels, the real and imaginary axes of $z$ are on the same $[-7,7]$ interval. } \label{f:roots}
\end{center}
\end{figure}

Roots of many Yablonskii-Vorob'ev polynomials $Q_{N}^{[m]}(z)$ have been plotted in the complex $z$ plane in \cite{Clarkson2003-II}. Critical to rogue waves in later sections, we replot such roots in Fig.~\ref{f:roots} for $2\le N\le 5$ and $1\le m\le N-1$. It is seen that these roots exhibit highly organized and symmetric structures, ranging from triangle to pentagon to heptagon and so on, depending on the value of $m$. In addition, all nonzero roots are simple, while the zero root may have higher multiplicity according to the above formula (\ref{QNmform}).

This root information of the polynomial $Q_{N}^{[m]}(z)$ turns out to be extremely important for the pattern analysis of rogue waves in Lemmas 1-3, since these roots are the ``DNA" of the wave pattern and codify its key information. Indeed, this intimate connection between rogue patterns and roots of the Yablonskii-Vorob'ev polynomials was first recognized in the NLS equation \cite{YangNLS2021}, where we showed that if the internal parameter $a_{2m+1}$ in the $N$-th order NLS rogue wave was large, then the geometric shape formed by the roots of $Q_{N}^{[m]}(z)$ gave the geometric structure of the NLS rogue pattern; each simple root led to a fundamental (Peregrine) rogue wave inside that pattern; and the zero root of multiplicity $N_0(N_0+1)/2$ resulted in a $N_0$-th order rogue wave at the center. In this article, we will show that similar connections between roots of $Q_{N}^{[m]}(z)$ and rogue patterns also hold for other integrable systems such as the ones presented earlier in this section.

\section{Analytical results on rogue patterns in the three integrable systems} \label{s:prediction}
It can be noticed that, similar to the NLS equation \cite{YangNLS2021}, rogue waves presented in the previous section for the three integrable systems also contain free internal complex parameters $\{a_3, a_5, \cdots, a_{2N-1}\}$. In this section, we will show that, when any of these internal parameters gets large, these rogue waves would also exhibit clear geometric patterns which are fully characterized by the root structures of the Yablonskii-Vorob'ev polynomial hierarchy, similar to the NLS equation~\cite{YangNLS2021}.

Suppose $a_{2m+1}$ is large, where $1\le m\le N-1$, and the value of $m$ is defined by the choice of $a_{2m+1}$
in the parameter set $\{a_3, a_5, \cdots, a_{2N-1}\}$ of rogue waves $u_N(x,t)$. Suppose also that the other parameters $\{a_3, a_5, \cdots, a_{2m-1}, a_{2m+3}, \cdots, a_{2N-1}\}$ are all $O(1)$. In addition, we introduce a quantity
\[ \label{defOmega}
\Omega\equiv \left(-\frac{2m+1}{2^{2m}}a_{2m+1}\right)^{\frac{1}{2m+1}}.
\]
Then, our analytical results on the large-$a_{2m+1}$ asymptotics of rogue waves in the GDNLS, Boussinesq and Manakov equations are summarized by the following three theorems.

\begin{quote}
\textbf{Theorem 1.}
\emph{For the GDNLS rogue wave $u_N(x,t)$ in Eq.~(\ref{BilinearTrans2}), when $a_{2m+1}$ is large and the other internal parameters $O(1)$, then}
\begin{enumerate}
\item \emph{far away from the origin, with $\sqrt{x^2+t^2}=O\left(|a_{2m+1}|^{1/(2m+1)}\right)$, this $u_N(x,t)$ asymptotically separates into $N_p$ fundamental rogue waves, where $N_p$ is given in Eq.~(\ref{Np}). These fundamental rogue waves are $\hat{u}_1(x-\hat{x}_{0}, t-\hat{t}_{0})\hspace{0.06cm} e^{{\rm i}(1-\gamma-\alpha) x-{\frac{\rm i}{2}}\left[\alpha^2+2(\gamma-2)\alpha+1-\gamma\right]t}$,
where
\[ \label{u1hatGDNLS}
|\hat{u}_{1}(x,t)|=\left|\frac{ \alpha (x+\alpha t)^2+(x-t)^2- {\rm{i}} (x+3 \alpha t) - \frac{3}{4} }{\alpha (x+\alpha t)^2+ (x-t)^2+{\rm{i}} (x+\alpha t)- 2 {\rm{i}} t+ \frac{1}{4}}\right|,
\]
and their positions $(\hat{x}_{0}, \hat{t}_{0})$ are given by
\begin{eqnarray}
&& \hat{x}_{0}= \frac{1}{\sqrt{\alpha}} \Re\left[ z_{0} \Omega \right]-\frac{\alpha-1}{\alpha} \Im\left[z_{0} \Omega \right], \label{x0Th1} \\
&& \hat{t}_{0}=\frac{1}{\alpha} \Im\left[ z_{0}\Omega \right],    \label{t0Th1}
\end{eqnarray}
where $z_0$ is any of the $N_p$ non-zero simple roots of $Q_{N}^{[m]}(z)$, and $(\Re, \Im)$ represent the real and imaginary parts of a complex number. The error of this fundamental rogue wave approximation is $O(|a_{2m+1}|^{-1/(2m+1)})$. Expressed mathematically, when $|a_{2m+1}|\gg 1$ and $\left[(x-\hat{x}_{0})^2+(t-\hat{t}_{0})^2\right]^{1/2}=O(1)$, we have the following solution asymptotics}
\[
u_{N}(x,t; a_{3}, a_{5}, \cdots, a_{2N-1}) = \hat{u}_1(x-\hat{x}_{0},t-\hat{t}_{0})\hspace{0.05cm} e^{{\rm i}(1-\gamma-\alpha) x-{\frac{\rm i}{2}}\left[\alpha^2+2(\gamma-2)\alpha+1-\gamma\right]t} + O\left(|a_{2m+1}|^{-1/(2m+1)}\right).
\]
\emph{When $(x,t)$ is not in the neighborhood of any of these $N_p$ fundamental waves, or $\sqrt{x^2+t^2}$ is larger than $O\left(|a_{2m+1}|^{1/(2m+1)}\right)$, then $u_N(x,t)$ asymptotically approaches the constant-amplitude background $e^{{\rm i}(1-\gamma-\alpha) x-{\frac{\rm i}{2}}\left[\alpha^2+2(\gamma-2)\alpha+1-\gamma\right]t}$ as $|a_{2m+1}|\to \infty$.}

\item
\emph{In the neighborhood of the origin, where $\sqrt{x^2+t^2}=O(1)$, $u_N(x,t)$ is approximately a lower $N_0$-th order rogue wave $u_{N_0}(x,t)$, where $N_0$ is given from $(N, m)$ as described in Sec.~\ref{sec:YV}, $0\le N_0\le N-2$, and $u_{N_0}(x,t)$ is given by Eq. (\ref{BilinearTrans2}) with its internal parameters $(a_3, a_5, \cdots, a_{2N_0-1})$ being the first $N_0-1$ values in the parameter set $\{a_3, a_5, \cdots, a_{2N-1}\}$ of the original rogue wave $u_N(x,t)$. The error of this lower-order rogue wave approximation $u_{N_0}(x,t)$ is $O(|a_{2m+1}|^{-1})$. Expressed mathematically, when $|a_{2m+1}|\gg 1$ and $\sqrt{x^2+t^2}=O(1)$,
\[ \label{ucenter1}
u_{N}(x,t; a_{3}, a_{5}, \cdots, a_{2N-1}) = u_{N_{0}}(x,t; a_{3}, a_{5}, \cdots, a_{2N_{0}-1})+ O\left(|a_{2m+1}|^{-1}\right).
\]
If $N_0=0$, then there will not be such a lower-order rogue wave in the neighborhood of the origin, and $u_N(x,t)$ asymptotically approaches the constant-amplitude background $e^{{\rm i}(1-\gamma-\alpha) x-{\frac{\rm i}{2}}\left[\alpha^2+2(\gamma-2)\alpha+1-\gamma\right]t}$ there as $|a_{2m+1}|\to \infty$.}
\end{enumerate}
\end{quote}

\begin{quote}
\textbf{Theorem 2.}
\emph{For the Boussinesq rogue wave $u_N(x,t)$ in Eq.~(\ref{Boussi_rogue}), when $a_{2m+1}$ is large and the other internal parameters $O(1)$, then}
\begin{enumerate}
\item \emph{far away from the origin, with $\sqrt{x^2+t^2}=O\left(|a_{2m+1}|^{1/(2m+1)}\right)$, this $u_N(x,t)$ asymptotically separates into $N_p$ fundamental rogue waves, where $N_p$ is given in Eq.~(\ref{Np}). These fundamental rogue waves are $u_1(x-\hat{x}_{0}, t-\hat{t}_{0})$, where
\[ \label{u1Boussi}
u_{1}(x,t)=2 \partial_{x}^2 \ln \left(x^2 + t^2+1\right),
\]
and their positions $(\hat{x}_{0}, \hat{t}_{0})$ are given by
\begin{eqnarray}
&& \hat{x}_{0}=\Re\left[\left(-2\rm{i} \sqrt{3}\right)z_{0}\Omega \right]+\Delta_B,  \label{x0Boussi}
\\
&& \hat{t}_{0}=\Im\left[\left(-2\rm{i} \sqrt{3}\right) z_{0} \Omega \right],     \label{t0Boussi}
\end{eqnarray}
where $z_0$ is any of the $N_p$ non-zero simple roots of $Q_{N}^{[m]}(z)$, and
\[ \label{DeltaB}
\Delta_B=-\frac{4}{3} (N-1).
\]
The error of this fundamental rogue wave approximation is $O(|a_{2m+1}|^{-1/(2m+1)})$. Expressed mathematically, when $|a_{2m+1}|\gg 1$ and $\left[(x-\hat{x}_{0})^2+(t-\hat{t}_{0})^2\right]^{1/2}=O(1)$, we have the following solution asymptotics
\[ \label{Th2uN}
u_{N}(x,t; a_{3}, a_{5}, \cdots, a_{2N-1}) = u_1(x-\hat{x}_{0},t-\hat{t}_{0}) + O\left(|a_{2m+1}|^{-1/(2m+1)}\right).
\]
When $(x,t)$ is not in the neighborhood of any of these $N_p$ fundamental waves, or $\sqrt{x^2+t^2}$ is larger than $O\left(|a_{2m+1}|^{1/(2m+1)}\right)$, then $u_N(x,t)$ asymptotically approaches the zero background as $|a_{2m+1}|\to \infty$.}

\item
\emph{In the neighborhood of the origin, where $\sqrt{x^2+t^2}=O(1)$, $u_N(x,t)$ is approximately a lower $N_0$-th order rogue wave $u_{N_0}(x,t)$, where $N_0$ is given from $(N, m)$ as described in Sec.~\ref{sec:YV}, $0\le N_0\le N-2$, and $u_{N_0}(x,t)$ is given by Eq. (\ref{Boussi_rogue}) with its new internal parameters $(\hat{a}_1, \hat{a}_3, \cdots, \hat{a}_{2N_0-1})$ related to the original parameters of $u_N(x,t)$ as
\[  \label{akBoussi}
\hat{a}_{2r-1}=a_{2r-1}+\left(N-N_{0}\right)s_{2r-1}, \quad r=1, 2, \cdots, N_0.
\]
The error of this lower-order rogue wave approximation $u_{N_0}(x,t)$ is $O(|a_{2m+1}|^{-1})$. Expressed mathematically, when $|a_{2m+1}|\gg 1$ and $\sqrt{x^2+t^2}=O(1)$,
\[ \label{ucenter2}
u_{N}(x,t; a_{1}, a_{3}, a_{5}, \cdots, a_{2N-1}) = u_{N_{0}}(x,t; \hat{a}_1, \hat{a}_3, \hat{a}_5, \cdots, \hat{a}_{2N_0-1})+ O\left(|a_{2m+1}|^{-1}\right).
\]
Note that while $a_1=0$ in the original rogue wave $u_N(x,t)$ (see Lemma 2), its counterpart $\hat{a}_1$ in the lower-order rogue wave $u_{N_{0}}(x,t)$ will not be zero. If $N_0=0$, then there will not be such a lower-order rogue wave in the neighborhood of the origin, and $u_N(x,t)$ asymptotically approaches the zero background there as $|a_{2m+1}|\to \infty$.}
\end{enumerate}
\end{quote}

\begin{quote}
\textbf{Theorem 3.}
\emph{For the Manakov rogue wave $[u_{1,N}(x,t), u_{2,N}(x,t)]$ in Eq.~(\ref{u1u2}), when $a_{2m+1}$ is large and the other internal parameters $O(1)$, then}
\begin{enumerate}
\item \emph{far away from the origin, with $\sqrt{x^2+t^2}=O\left(|a_{2m+1}|^{1/(2m+1)}\right)$, this $[u_{1,N}(x,t), u_{2,N}(x,t)]$ asymptotically separates into $N_p$ fundamental rogue waves, where $N_p$ is given in Eq.~(\ref{Np}). These fundamental rogue waves are $[\hat{u}_{1,1}(x-\hat{x}_{0}, t-\hat{t}_{0})\hspace{0.06cm} e^{{\rm{i}} (k_1x+\omega_{1} t)}, \hspace{0.07cm} \hat{u}_{2,1}(x-\hat{x}_{0}, t-\hat{t}_{0})\hspace{0.06cm} e^{{\rm{i}} (k_{2}x + \omega_{2} t)}]$, where}
\[\label{u11Mana}
\hat{u}_{1,1}(x,t)=\rho_1 \frac{\left[ p_{1} x+ 2 p_{0} p_{1}\left(  \textrm{i}t  \right) + \theta_{1}\right] \left[ p_{1}^* x- 2 p_{0}^* p_{1}^*\left(\textrm{i}t  \right) - \theta_{1}^* \right]+\zeta_{0}}{\left| p_{1} x+ 2 p_{0} p_{1}\left(  \textrm{i}t  \right) \right|^2+\zeta_{0}},
\]
\[ \label{u21Mana}
\hat{u}_{2,1}(x,t)=\rho_2 \frac{\left[ p_{1} x+ 2 p_{0} p_{1}\left(  \textrm{i}t  \right) + \lambda_{1}\right] \left[ p_{1}^* x- 2 p_{0}^* p_{1}^*\left(\textrm{i}t  \right)  -  \lambda_{1}^* \right]+\zeta_{0}}{\left| p_{1} x+ 2 p_{0} p_{1}\left(  \textrm{i}t  \right) \right|^2+\zeta_{0}},
\]
\[
\theta_{1}=\frac{p_{1}}{p_{0}-\textrm{i} k_{1}}, \ \lambda_{1}=\frac{p_{1}}{p_{0}-\textrm{i} k_{2}}, \ \zeta_{0} =\frac{|p_{1}|^2}{(p_{0}+p_{0}^*)^2},
\]
\emph{and their positions $(\hat{x}_{0}, \hat{t}_{0})$ are given by
\begin{eqnarray}
&& \hat{x}_{0}= \frac{1}{\Re(p_0)}\Re\left[ \frac{z_{0}\Omega}{p_{1}}p_{0}^* \right]+\Delta_{1M},  \label{x0Mana} \\
&& \hat{t}_{0}=\frac{1}{2 \Re(p_0)} \Im\left[ \frac{z_{0}\Omega}{p_{1}} \right]+\Delta_{2M},  \label{t0Mana}
\end{eqnarray}
where $z_0$ is any of the $N_p$ non-zero simple roots of $Q_{N}^{[m]}(z)$, and
\begin{equation} \label{DeltaM1M2}
\Delta_{1M}=-\frac{1}{\Re(p_0)}\Re\left[ \frac{(N-1)s_1}{p_1}p_0^* \right], \quad
\Delta_{2M}=-\frac{1}{2 \Re(p_0)} \Im\left[\frac{(N-1)s_1}{p_1} \right].
\end{equation}
The error of this fundamental rogue wave approximation is $O(|a_{2m+1}|^{-1/(2m+1)})$. Expressed mathematically, when $|a_{2m+1}|\gg 1$ and $\left[(x-\hat{x}_{0})^2+(t-\hat{t}_{0})^2\right]^{1/2}=O(1)$, we have the following solution asymptotics
\begin{eqnarray}
&& u_{1,N}(x,t; a_{3}, a_{5}, \cdots, a_{2N-1}) = \hat{u}_{1,1}(x-\hat{x}_{0},t-\hat{t}_{0}) \hspace{0.06cm} e^{{\rm{i}} (k_1x+\omega_{1} t)} + O\left(|a_{2m+1}|^{-1/(2m+1)}\right),   \label{u1NTh3} \\
&& u_{2,N}(x,t; a_{3}, a_{5}, \cdots, a_{2N-1}) = \hat{u}_{2,1}(x-\hat{x}_{0},t-\hat{t}_{0}) \hspace{0.06cm} e^{{\rm{i}} (k_2x+\omega_{2} t)} + O\left(|a_{2m+1}|^{-1/(2m+1)}\right).   \label{u2NTh3}
\end{eqnarray}
When $(x,t)$ is not in the neighborhood of any of these $N_p$ fundamental waves, or $\sqrt{x^2+t^2}$ is larger than $O\left(|a_{2m+1}|^{1/(2m+1)}\right)$, then $[u_{1,N}(x,t), u_{2,N}(x,t)]$ asymptotically approaches the constant-amplitude background $[\rho_{1}  e^{{\rm{i}} (k_{1}x + \omega_{1} t)}, \rho_{2}  e^{{\rm{i}} (k_{2}x + \omega_{2} t)}]$ as $|a_{2m+1}|\to \infty$.}

\item
\emph{In the neighborhood of the origin, where $\sqrt{x^2+t^2}=O(1)$, $[u_{1,N}(x,t), u_{2,N}(x,t)]$ is approximately a lower $N_0$-th order rogue wave $[u_{1,N_0}(x,t), u_{2,N_0}(x,t)]$, where $N_0$ is given from $(N, m)$ as described in Sec.~\ref{sec:YV}, $0\le N_0\le N-2$, and $[u_{1,N_0}(x,t), u_{2,N_0}(x,t)]$ is given by Eq.~(\ref{u1u2}) with its new internal parameters $(\hat{a}_1, \hat{a}_3, \cdots, \hat{a}_{2N_0-1})$  related to the original parameters of $[u_{1,N}(x,t), u_{2,N}(x,t)]$ as
\[ \label{akMana}
\hat{a}_{2r-1}=a_{2r-1}+\left(N-N_{0}\right)s_{2r-1}, \quad r=1, 2, \cdots, N_0.
\]
The error of this lower-order rogue wave approximation is $O(|a_{2m+1}|^{-1})$. Expressed mathematically, when $|a_{2m+1}|\gg 1$ and $\sqrt{x^2+t^2}=O(1)$,
\begin{eqnarray}
&& u_{1,N}(x,t; a_{1}, a_{3}, a_{5}, \cdots, a_{2N-1}) = u_{1,N_{0}}(x,t; \hat{a}_1, \hat{a}_3, \hat{a}_5, \cdots, \hat{a}_{2N_0-1}) + O\left(|a_{2m+1}|^{-1}\right), \label{u1NTh3b} \\
&& u_{2,N}(x,t; a_{1}, a_{3}, a_{5}, \cdots, a_{2N-1}) = u_{2,N_{0}}(x,t; \hat{a}_1, \hat{a}_3, \hat{a}_5, \cdots, \hat{a}_{2N_0-1}) + O\left(|a_{2m+1}|^{-1}\right).  \label{u2NTh3b}
\end{eqnarray}
Note that $a_1=0$ in the original rogue wave $[u_{1,N}(x,t), u_{2,N}(x,t)]$, but its counterpart $\hat{a}_1$ in the lower-order rogue wave $[u_{1,N_0}(x,t), u_{2,N_0}(x,t)]$ will not be zero. If $N_0=0$, then there will not be such a lower-order rogue wave in the neighborhood of the origin, and $[u_{1,N}(x,t), u_{2,N}(x,t)]$ asymptotically approaches the constant-amplitude background $[\rho_{1}  e^{{\rm{i}} (k_{1}x + \omega_{1} t)}, \rho_{2}  e^{{\rm{i}} (k_{2}x + \omega_{2} t)}]$ there as $|a_{2m+1}|\to \infty$.}
\end{enumerate}
\end{quote}

One can clearly see from these theorems that rogue patterns under the large-parameter limit in these three integrable systems have a lot in common. In all cases, the center of the wave pattern is a lower-order rogue wave, and the order of this center rogue wave depends only on the order of the original rogue wave and the index of the large internal parameter. Away from the center, the rogue wave comprises fundamental rogue waves, whose number is equal to the number of nonzero simple roots of the Yablonskii-Vorob'ev polynomial $Q_{N}^{[m]}(z)$, and whose $(x, t)$ locations are linearly dependent on the real and imaginary parts of these nonzero simple roots. To put it mathematically, the location $(\hat{x}_{0}, \hat{t}_{0})$ of each fundamental rogue wave inside the rogue structure is given by the real and imaginary parts of each nonzero simple root $z_0$ of $Q_{N}^{[m]}(z)$ through a linear transformation
\[ \label{x0t0B}
\left[\begin{array}{c} \hat{x}_{0} \\ \hat{t}_{0} \end{array}\right]=
\textbf{B}\left[\begin{array}{c} \Re(z_0) \\ \Im(z_0) \end{array}\right]+
\left[\begin{array}{c} \Delta_{1} \\ \Delta_{2} \end{array}\right],
\]
where $\textbf{B}$ is a constant matrix and $(\Delta_{1}, \Delta_{2})$ a constant vector. For the GDNLS equations (\ref{GDNLS}), it is easy to see from Eqs.~(\ref{x0Th1})-(\ref{t0Th1}) of Theorem~1 that
\[ \label{BGDNLS}
\textbf{B}=\left[\begin{array}{cc} \frac{1}{\sqrt{\alpha}}\Re(\Omega)-\frac{\alpha-1}{\alpha}\Im(\Omega) & -\frac{1}{\sqrt{\alpha}}\Im(\Omega)-\frac{\alpha-1}{\alpha}\Re(\Omega)  \\ \frac{1}{\alpha}\Im(\Omega) & \frac{1}{\alpha}\Re(\Omega) \end{array}\right], \qquad \left[\begin{array}{c} \Delta_{1} \\ \Delta_{2} \end{array}\right]=\left[\begin{array}{c} 0 \\ 0 \end{array}\right].
\]
For the Boussinesq equation (\ref{BoussiEq}), Eqs.~(\ref{x0Boussi})-(\ref{t0Boussi}) of Theorem~2 give
\[ \label{BBoussi}
\textbf{B}=2\sqrt{3}\left[\begin{array}{cc} \Im(\Omega) & \Re(\Omega) \\ -\Re(\Omega) & \Im(\Omega) \end{array}\right], \qquad \left[\begin{array}{c} \Delta_{1} \\ \Delta_{2} \end{array}\right]=\left[\begin{array}{c} \Delta_{B} \\ 0 \end{array}\right],
\]
where $\Delta_B$ is provided by Eq.~(\ref{DeltaB}). For the Manakov system (\ref{ManakModel1})-(\ref{ManakModel2}), Eqs.~(\ref{x0Mana})-(\ref{t0Mana}) of Theorem~3 give
\[ \label{BMana}
\textbf{B}=\frac{1}{\Re(p_0)}
\left[\begin{array}{cc} \Re\left(\frac{p_0^*\Omega}{p_1}\right) & -\Im\left(\frac{p_0^*\Omega}{p_1}\right) \\ \frac{1}{2}\Im\left(\frac{\Omega}{p_1}\right) & \frac{1}{2}\Re\left(\frac{\Omega}{p_1}\right) \end{array}\right], \qquad \left[\begin{array}{c} \Delta_{1} \\ \Delta_{2} \end{array}\right]=\left[\begin{array}{c} \Delta_{1M} \\ \Delta_{2M} \end{array}\right],
\]
where $(\Delta_{1M}, \Delta_{2M})$ are provided by Eq.~(\ref{DeltaM1M2}). It is important to notice that the constant matrix $\textbf{B}$ and the constant vector $(\Delta_{1}, \Delta_{2})$ are both independent of the root $z_0$, which is why (\ref{x0t0B}) is a linear transformation from the $z$-plane to the $(x,t)$ plane.
This linear transformation means that the whole rogue pattern formed by fundamental rogue waves in the $(x,t)$ plane is just a linear transformation matrix $\textbf{B}$ applied to the root structure of the Yablonskii-Vorob'ev polynomial $Q_{N}^{[m]}(z)$ in the complex $z$ plane, plus a constant position shift $(\Delta_{1}, \Delta_{2})$. For the GDNLS equations with $\alpha=1$, $\textbf{B}/|\Omega|$ is a rotation matrix (since it is orthogonal with determinant 1). For the Boussinesq equation, $\textbf{B}/\left(2\sqrt{3}\hspace{0.06cm} |\Omega|\right)$ is a rotation matrix. For the Manakov system with $p_0=1/2$, $\textbf{B}/(|p_1|^{-1}|\Omega|)$ is a rotation matrix.
In these cases, the $\textbf{B}$ part of the transformation is just a combination of rotation and dilation to the root structure. But for the GDNLS equations with $\alpha\ne 1$ and the Manakov system with $p_0\ne 1/2$, the $\textbf{B}$ part of the transformation would also involve other actions such as shear and stretch to the root structure.

\begin{figure}[htb]
\begin{center}
\vspace{-2.5cm}
\hspace{-4.5cm}
\includegraphics[scale=0.6, bb=0 0 385 567]{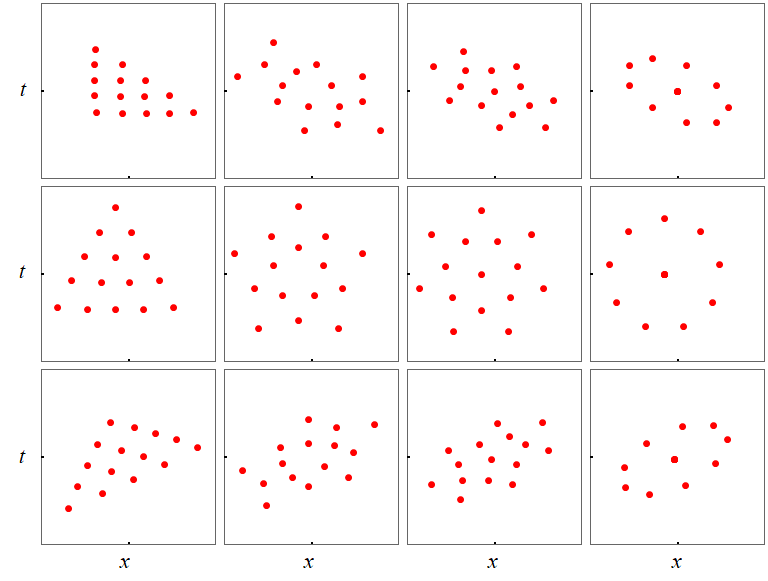}\\
\caption{The $(x, t)$ plane images of the linear transformation (\ref{x0t0B}) on the root structures of $Q_{5}^{[m]}(z)$ with $1\le m\le 4$, for the three integrable systems with the respective large internal parameter $a_{2m+1}$ and other system parameters provided by Eqs.~(\ref{paraGDNLS})-(\ref{paraMana}). Upper row: images for the GDNLS equations, where $-13\le x, t \le 13$. Middle row: images for the Boussinesq equation, where $-35 \le x, t \le 35$. Lower row: images for the Manakov system, where $-28 \le x, t \le 28$.} \label{images}
\end{center}
\end{figure}

To get a visual impression of the linear transformation (\ref{x0t0B}), we plot in Fig.~\ref{images} its effects on the root structures of $Q_{5}^{[m]}(z)$ with $1\le m\le 4$ for the underlying three integrable systems. The reader is reminded that these root structures of $Q_{5}^{[m]}(z)$ have been plotted in the bottom row of Fig.~\ref{f:roots}. In the top row of Fig.~\ref{images}, the images of the transformation (\ref{x0t0B}) on these root structures are displayed for the GDNLS equations with the background wavenumber $\alpha=16/9$ and the large internal parameter $a_{2m+1}$ respectively as
\[ \label{paraGDNLS}
(a_3, a_5, a_7, a_9)=(-15 \textrm{i}, -250 \textrm{i}, -1000 \textrm{i}, -3000 \textrm{i}).
\]
The middle row of Fig.~\ref{images} are images of this transformation for the Boussinesq equation with the large parameter $a_{2m+1}$ respectively as
\[ \label{paraBoussi}
(a_3, a_5, a_7, a_9)=(-5, 20, -80, 200).
\]
In the bottom row of Fig.~\ref{images}, images of the transformation for the Manakov equations are illustrated under system and background parameter choices of
\[\label{ManaparFig1}
\epsilon_1=1, \quad \epsilon_2=1, \quad k_1=\frac{1}{2}, \quad k_2=-\frac{1}{2}, \quad \rho_1=1, \quad \rho_2=1, \quad p_{0}=\frac{1}{2} \sqrt{1+2 \rm{i} },
\]
with the large parameter $a_{2m+1}$ respectively as
\[ \label{paraMana}
(a_3, a_5, a_7, a_9)=(40, 400, 3000, 20000).
\]

By comparing these images to the original root structures at the bottom row of Fig.~\ref{f:roots}, we can see that in the Boussinesq case, the images (middle row of Fig.~\ref{images}) have the same shapes of the root structures but with different orientations. To obtain these orientation angles, we notice from Eqs.~(\ref{x0Boussi})-(\ref{t0Boussi}) of Theorem~2 that
\[
\hat{x}_{0}+{\rm{i}}\hat{t}_{0}=\left(-2\rm{i} \sqrt{3}\right)z_{0}\Omega+\Delta_B.
\]
Thus, the orientation angle of the image is equal to that of the root structure plus arg($\Omega$) and minus $\pi/2$. Notice from the definition (\ref{defOmega}) of $\Omega$ that arg($\Omega$) is controlled by the phase of $a_{2m+1}$. Then, using the phase of $a_{2m+1}$, we can quickly predict the orientation angle of the image. For example, with $a_3=-5$ in the first panel of the middle row of Fig.~\ref{images}, arg($\Omega$)=0. Then, the orientation of this image should be that of the root structure rotated close-wise by $\pi/2$, which is clearly the case. Orientation angles of other panels in the middle row of Fig.~\ref{images} can be predicted similarly.

In the GDNLS and Manakov cases, however, the images (top and bottom rows of Fig.~\ref{images}) look different from the root structures, because stretch and shear are clearly seen in addition to changes in orientation. As we have mentioned before, this stretch and shear is induced by the underlying linear transformation matrix $\textbf{B}$ not being a rotation matrix, since in Fig.~\ref{images}, $\alpha\ne 1$ in the GDNLS images and $p_0\ne 1/2$ in the Manakov images. In these cases, the amounts of stretch, shear and orientation shift are all dictated by details of the $\textbf{B}$ matrix.

\section{Predicted rogue patterns and comparison with true ones in the three integrable systems} \label{sec:comparison}

In this section, we illustrate our analytical predictions of rogue patterns in Theorems 1-3 and compare them to true rogue patterns in the underlying three integrable systems.

\subsection{Prediction and comparison for the GDNLS equations}

\begin{figure}[htb]
\begin{center}
\includegraphics[scale=1.0, bb=0 0 385 397]{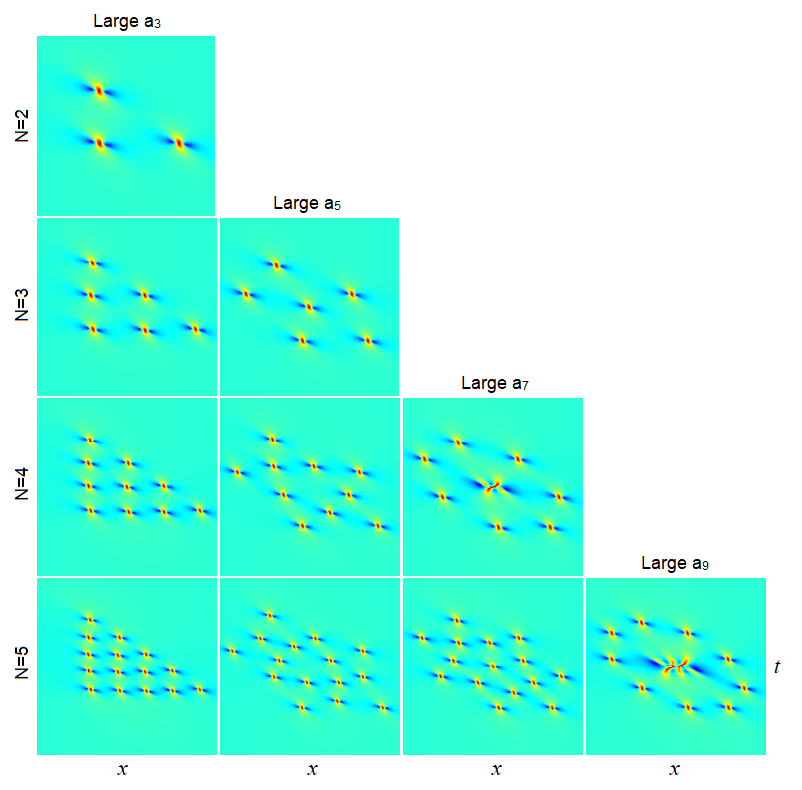}
\caption{Predicted GDNLS rogue patterns $|u_N(x,t)|$ for the orders $2\le N\le 5$ and the large parameter $a_{2m+1}$ from $m=1$ to $N-1$ [the background wavenumber parameter in Eq.~(\ref{BoundaryCond1}) is chosen as $\alpha=16/9$]. For each panel, the large parameter $a_{2m+1}$ in the rogue wave is displayed in Table~1, with the other internal parameters set as zero. The center of each panel is always the origin $x=t=0$, but the $(x,t)$ intervals differ slightly from panel to panel. For instance, in the bottom row, the left-most panel has $-12 \le x, t \le 12$, and the right-most panel has $-10 \le x, t \le 10$.
} \label{f:GDNLS1}
\end{center}
\end{figure}

First, we present the prediction and comparison for the GDNLS equations (\ref{GDNLS}). According to Theorem~1, our predictions for their rogue patterns under a large internal parameter $a_{2m+1}$ can be assembled into a single formula
\[\label{approximaRWs1}
\left|u_{N}(x,t)\right|\thickapprox \left|u_{N_{0}}(x,t)\right| + \sum _{k=1}^{N_{p}}  \left(\left| \hat{u}_{1}(x-\hat{x}_{0}^{(k)},t-\hat{t}_{0}^{(k)})\right| -1 \right),
\]
where $N_0$ is given from $(N, m)$ as described in Sec.~\ref{sec:YV}, $u_{N_{0}}(x,t)$ is the lower-order rogue wave at the center whose internal parameters $(a_3, a_5, \cdots a_{2N_0-1})$ are inherited directly from those of the original rogue wave $u_N(x,t)$, the function $\hat{u}_1(x,t)$ is the fundamental rogue wave given in Eq.~(\ref{u1hatGDNLS}), with its position $(\hat{x}_{0}^{(k)}, \hat{t}_{0}^{(k)})$ given by Eqs.~(\ref{x0Th1})-(\ref{t0Th1}) for every one of the nonzero simple roots $z_0^{(k)}$ of $Q_{N}^{[m]}(z)$, and $N_p$ is the number of such fundamental rogue waves whose value is given by Eq.~(\ref{Np}). Then, when we choose $\alpha=16/9$ in the background (\ref{BoundaryCond1}), the above amplitude approximations (\ref{approximaRWs1}) for $2\le N\le 5$, with the large internal parameter $a_{2m+1}$ ranging from $m=1$ to $N-1$ and its value taken as in Table 1, and with the other internal parameters taken as zero, are displayed in Fig.~\ref{f:GDNLS1}. Notice that the predicted patterns are stretched triangles in the first column (for large $a_3$), stretched pentagons in the second column (for large $a_5$), stretched heptagons in the third column (for large $a_7$), and so on. To understand how these predicted patterns come about, let us consider these patterns in the bottom row of this figure (where $N=5$). Notice that the large internal parameters in this row of rogue patterns are the same as those in Eq.~(\ref{paraGDNLS}) for the top (GDNLS) row of the linear-transformation image figure \ref{images}. Thus, according to our Theorem 1, that top row of Fig.~\ref{images} gives the predicted locations of fundamental rogue waves which form the rogue pattern. When we flesh out those top panels by replacing each non-center dot with a GDNLS fundamental rogue wave and replacing the center dot by a lower $N_0$-th order rogue wave, then we get the bottom row of Fig.~\ref{f:GDNLS1}.  Thus, the predicted rogue patterns in the bottom row of Fig.~\ref{f:GDNLS1} are inherited directly from the top row of Fig.~\ref{images}, i.e., from images of the linear transformation (\ref{x0t0B}) on the root structure of the Yablonskii-Vorob'ev polynomials. The other predicted rogue patterns in Fig.~\ref{f:GDNLS1} can be understood similarly through the linear transformations (\ref{x0t0B}).

\begin{table}[h]
\caption{Value of the large parameter for GDNLS rogue waves in Fig.~\ref{f:GDNLS1} with $\alpha=16/9$}
\begin{center}
  \begin{tabular}{ | c | c | c|  c | c|  c | c|}  \hline
         $N$ &              $a_{3}$   & $a_{5}$    & $a_{7}$   & $a_{9}$       \\ \hline
          2 &               $ -30 \textrm{i}$  &    &   &     \\ \hline
          3 &             $ -28 \textrm{i}$  &  $-500 \textrm{i}$ &     &    \\ \hline
          4 &              $ -20 \textrm{i}$  &  $-300 \textrm{i}$ & $-1500\textrm{i}$  &      \\ \hline
          5 &               $ -15 \textrm{i}$  &  $-250 \textrm{i}$  & $-1000 \textrm{i}$ &$ -3000 \textrm{i}$   \\ \hline
  \end{tabular}
\end{center}
\end{table}

\begin{figure}[htb]
\begin{center}
\includegraphics[scale=0.5, bb=340 0 385 227]{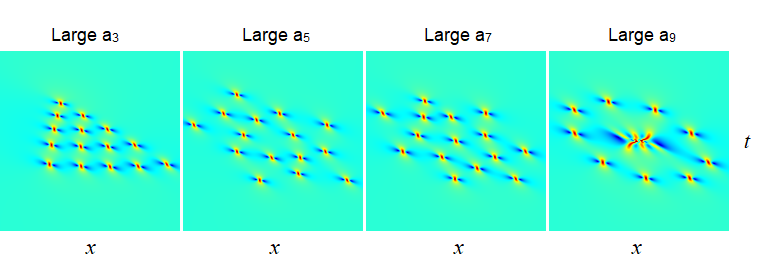}
\caption{True GDNLS rogue patterns $|u_N(x,t)|$ for $N=5$. The $\alpha$ value, internal parameters and $(x,t)$ intervals for these true solutions are identical to those in the theoretically predicted patterns shown in the bottom row of Fig.~\ref{f:GDNLS1}. }  \label{f:GDNLS2}
\end{center}
\end{figure}

Now, we compare these predicted rogue patterns with true ones. For brevity, we only show this comparison for $N=5$. Under identical $\alpha$ and internal parameter choices and identical $(x,t)$ intervals as in the bottom row of Fig.~\ref{f:GDNLS1}, true rogue patterns are displayed in Fig.~\ref{f:GDNLS2}. It is seen that the true rogue patterns are almost indistinguishable from the predictions on all aspects, from the locations of individual fundamental rogue waves, to the overall shapes formed by these fundamental waves, and to the fine details of the lower-order rogue waves at the center. Similar agreements hold for the other panels of Fig.~\ref{f:GDNLS1} as well. These apparent visual agreements testify to the power and accuracy of our theoretical predictions.

Our Theorem 1 also states quantitatively that the error of the predicted solution in the neighborhood of each fundamental rogue wave away from the origin is $O(|a_{2m+1}|^{-1/(2m+1)})$, while that error in the neighborhood of the origin is $O(|a_{2m+1}|^{-1})$. We have numerically verified the orders of these errors similar to what we did earlier for the NLS equation in Ref.~\cite{YangNLS2021}, but details will not be shown for the sake of brevity.

\subsection{Prediction and comparison for the Boussinesq equation}

\begin{figure}[htb]
\begin{center}
\includegraphics[scale=1.0, bb=90 0 325 377]{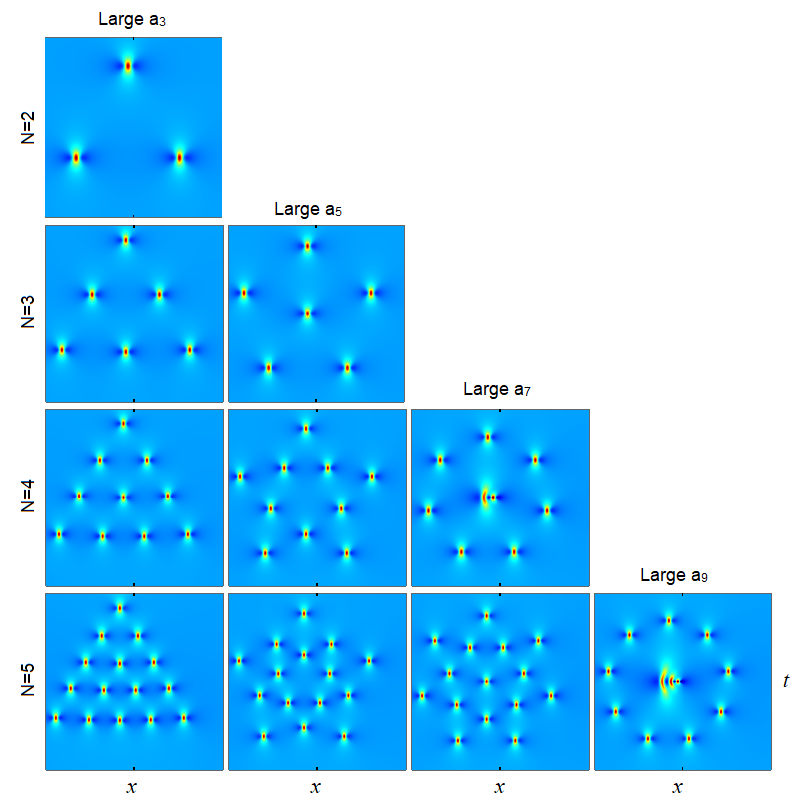}
\caption{Predicted Boussinesq rogue patterns $u_N(x,t)$ for the orders $2\le N\le 5$ and the large parameter $a_{2m+1}$ from $m=1$ to $N-1$. For each panel, the large parameter $a_{2m+1}$ in the rogue wave is displayed in Table~2, with the other internal parameters set as zero. The center of each panel is always the origin $x=t=0$, but the $(x,t)$ intervals differ slightly from panel to panel. For instance, in the bottom row, the first two panels have $-32.25 \le x, t \le 32.25$ and $-35.25 \le x, t \le 35.25$ respectively. }  \label{f:Boussi1}
\end{center}
\end{figure}

Next, we present prediction and comparison of rogue patterns for the Boussinesq equation. In this case, our prediction from Theorem~2 for the Boussinesq rogue pattern under a large internal parameter $a_{2m+1}$ can be assembled into the formula
\[\label{approximaRWs2}
u_{N}(x,t) \thickapprox  u_{N_{0}}(x,t) + \sum _{k=1}^{N_{p}} u_{1}\left(x-\hat{x}_{0}^{(k)}, t-\hat{t}_{0}^{(k)}\right),
\]
where $N_0$ is as given in Sec.~\ref{sec:YV}, $u_{N_{0}}(x,t)$ is the lower-order rogue wave at the center whose new internal parameters $(\hat{a}_1, \hat{a}_3, \cdots, \hat{a}_{2N_0-1})$ are given by Eq.~(\ref{akBoussi}), the function $u_1(x,t)$ is the fundamental Boussinesq rogue wave given in Eq.~(\ref{u1Boussi}), with its position $(\hat{x}_{0}^{(k)}, \hat{t}_{0}^{(k)})$ given by Eqs.~(\ref{x0Boussi})-(\ref{t0Boussi}) for every one of the nonzero simple roots $z_0^{(k)}$ of $Q_{N}^{[m]}(z)$, and $N_p$ is the number of such fundamental rogue waves whose value is given by Eq.~(\ref{Np}). When the large internal parameter $a_{2m+1}$ is selected as in Table 2 for $2\le N\le 5$ and $1\le m\le N-1$, with the other internal parameters set as zero, the predicted rogue solutions (\ref{approximaRWs2}) are illustrated in Fig.~\ref{f:Boussi1}. Notice that for the Boussinesq equation where the linear transformation matrix $\textbf{B}$ is as given in Eq.~(\ref{BBoussi}), $\textbf{B}/\left(2\sqrt{3}\hspace{0.06cm} |\Omega|\right)$ is a rotation matrix. Thus, the predicted rogue patterns in this figure can be obtained from the root structures of the Yablonskii-Vorob'ev polynomials through the actions of rotation, dilation and uniform shift but without stretch and shear --- a fact that is obvious by visually comparing these rogue patterns with the root structures of the Yablonskii-Vorob'ev polynomials displayed in Fig.~\ref{f:roots}. Like the GDNLS case before, the predicted rogue patterns in the bottom row of Fig.~\ref{f:Boussi1} (for $N=5$), whose large internal parameter values are the same as those in the earlier Eq.~(\ref{paraBoussi}), are also the flesh-out of the images of linear transformation (\ref{x0t0B}) on the Yablonskii-Vorob'ev root structures (those images were shown in the middle row of Fig.~\ref{images}). Notice that the current linear transformation involves a uniform $x$-position shift [see Eqs.~(\ref{DeltaB}) and (\ref{BBoussi})], which results in an overall $x$-shift of $\Delta_B=-4(N-1)/3$ to the whole rogue pattern. This shift is quite visible in Fig.~\ref{f:Boussi1}, especially for higher order rogue waves, because the amount $\Delta_B$ of this shift increases with $N$. But since the Boussinesq equation is translation-invariant in $(x,t)$, this overall $x$-shift does not affect the shape of the rogue pattern.

\begin{table}[h]
\caption{Value of the large parameter for Boussinesq rogue waves in Fig.~\ref{f:Boussi1}}
\begin{center}
  \begin{tabular}{ | c | c | c|  c | c|  c | }  \hline
         $N$ &              $a_{3}$   & $a_{5}$    & $a_{7}$   & $a_{9}$     \\ \hline
          2 &               $ -20 $  &    &   &        \\ \hline
          3 &             $ -16 $  &  $ 150 $ &     &       \\ \hline
          4 &              $ -10  $  &  $50 $ & $-100 $ &        \\ \hline
          5 &               $ -5  $  &  $20  $ & $-80 $ &$ 200  $      \\ \hline
  \end{tabular}
\end{center}
\end{table}

\begin{figure}[htb]
\begin{center}
\includegraphics[scale=0.5, bb=340 0 385 227]{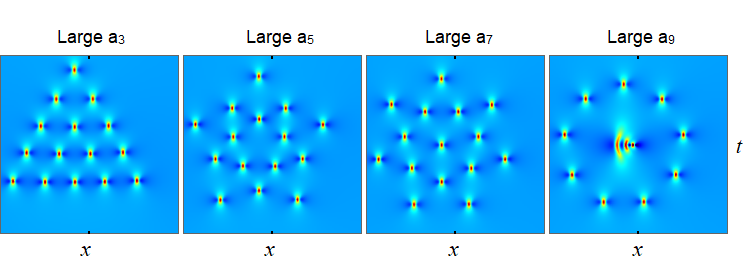}
\caption{True Boussinesq rogue patterns $u_N(x,t)$ for $N=5$. The internal parameters and $(x,t)$ intervals for these true solutions are identical to those in the theoretically predicted patterns shown in the bottom row of  Fig.~\ref{f:Boussi1}.  } \label{f:Boussi2}
\end{center}
\end{figure}

Now, we compare these predicted Boussinesq rogue patterns of Fig.~\ref{f:Boussi1} to the true solutions. Again, for brevity, we only do the comparison for the fifth-order rogue waves ($N=5$). Under the same internal parameter choices and $(x,t)$ intervals as those in the bottom row of Fig.~\ref{f:Boussi1}, the true Boussinesq rogue patterns are displayed in Fig.~\ref{f:Boussi2}. It is seen that again, the true rogue patterns are visually indistinguishable from our theoretical predictions on all aspects, from locations of fundamental rogue waves away from the center, to fine details of lower-order rogue waves at the center, and to the amounts of $x$-position shifts to the whole structure. Order of accuracy of our predictions as stated in Theorem~2 has also been confirmed numerically, with the details omitted.

\subsection{Prediction and comparison for the Manakov system}

\begin{figure}[htb]
\begin{center}
\includegraphics[scale=0.80, bb=200 0 385 330]{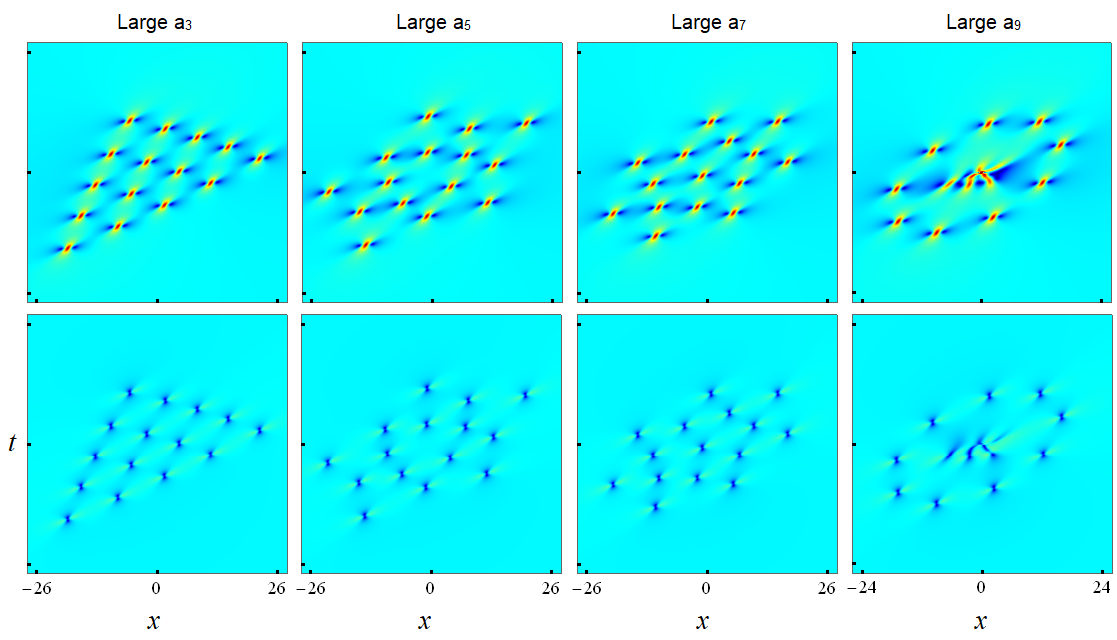}
\caption{Predicted patterns of fifth-order rogue waves in the Manakov equations under system parameters (\ref{ManaparFig1}) (upper row: $|u_{1,N}|$; lower row: $|u_{2,N}|$). The large internal parameter is as shown in Eq.~(\ref{paraMana}), i.e., $a_3=40$ in the first column, $a_5=400$ in the second column, $a_7=3000$ in the third column, $a_9=20000$ in the fourth column, and the other internal parameters are set as zero. }  \label{f:Mana1}
\end{center}
\end{figure}

Now, we consider the Manakov case. Our prediction from Theorem~3 for Manakov rogue patterns can be assembled into the formulae
\begin{eqnarray}
&& \left| u_{1,N}(x,t) \right| \thickapprox  \left| u_{1,N_{0}}(x,t) \right| + \sum _{k=1}^{N_{p}}  \left(\left| \hat{u}_{1,1}(x-\hat{x}_{0}^{(k)},t-\hat{t}_{0}^{(k)})\right| -\rho_1 \right),   \label{approximaRWs3a}  \\
&& \left| u_{2,N}(x,t) \right| \thickapprox  \left| u_{2,N_{0}}(x,t) \right| + \sum _{k=1}^{N_{p}}  \left(\left| \hat{u}_{2,1}(x-\hat{x}_{0}^{(k)},t-\hat{t}_{0}^{(k)})\right| -\rho_2 \right),   \label{approximaRWs3b}
\end{eqnarray}
where $N_0$ is as given in Sec.~\ref{sec:YV}, $[u_{1,N_{0}}(x,t), u_{2,N_{0}}(x,t)]$ is the lower-order rogue wave at the center whose new internal parameters $(\hat{a}_1, \hat{a}_3, \cdots, \hat{a}_{2N_0-1})$ are given by Eq.~(\ref{akMana}), the functions $[\hat{u}_{1,1}(x,t), \hat{u}_{2,1}(x,t)]$ are the fundamental Manakov rogue wave given in Eqs.~(\ref{u11Mana})-(\ref{u21Mana}), with their positions $(\hat{x}_{0}^{(k)}, \hat{t}_{0}^{(k)})$ given by Eqs.~(\ref{x0Mana})-(\ref{t0Mana}) for every one of the nonzero simple roots $z_0^{(k)}$ of $Q_{N}^{[m]}(z)$, and $N_p$ is the number of such fundamental rogue waves whose value is given by Eq.~(\ref{Np}). Since the Manakov waves have two components, to show both components, we will make the prediction and comparison for $N=5$ only. With the system and background parameters chosen as in Eq.~(\ref{ManaparFig1}), and with the large internal parameter $a_{2m+1}$ taken as in Eq.~(\ref{paraMana}) respectively, and the other internal parameters set as zero, the predicted rogue solutions (\ref{approximaRWs3a})-(\ref{approximaRWs3b}) are plotted in Fig.~\ref{f:Mana1}. These predicted rogue patterns are a flesh-out of the bottom panels in Fig.~\ref{images}, which are images of the transformation (\ref{x0t0B}) on the Yablonskii-Vorob'ev root structures for the Manakov system under the same large-parameter values. In the current case, the linear transformation matrix $\textbf{B}$ in Eq.~(\ref{BMana}) induces stretch and shear to the Yablonskii-Vorob'ev root structure, and this skewed pattern is evident in Fig.~\ref{f:Mana1}. Another feature in the present Manakov case is that the predicted rogue patterns contain uniform shifts in both $x$ and $t$, since $\Delta_{1}$ and $\Delta_{2}$ in Eq.~(\ref{BMana}) are both nonzero. However, under our choices of system and background parameters (\ref{ManaparFig1}) and for $N=5$, we find from Eq.~(\ref{DeltaM1M2}) that
$\Delta_{1}\approx -1.07433$ and $\Delta_{2}\approx -0.68328$, which are both relatively small. Thus, these $(x,t)$ shifts to the rogue patterns in Fig.~\ref{f:Mana1} are not as pronounced as those for the Boussinesq equation in Fig.~\ref{f:Boussi1}.

\begin{figure}[htb]
\begin{center}
\includegraphics[scale=0.80, bb=200 0 385 330]{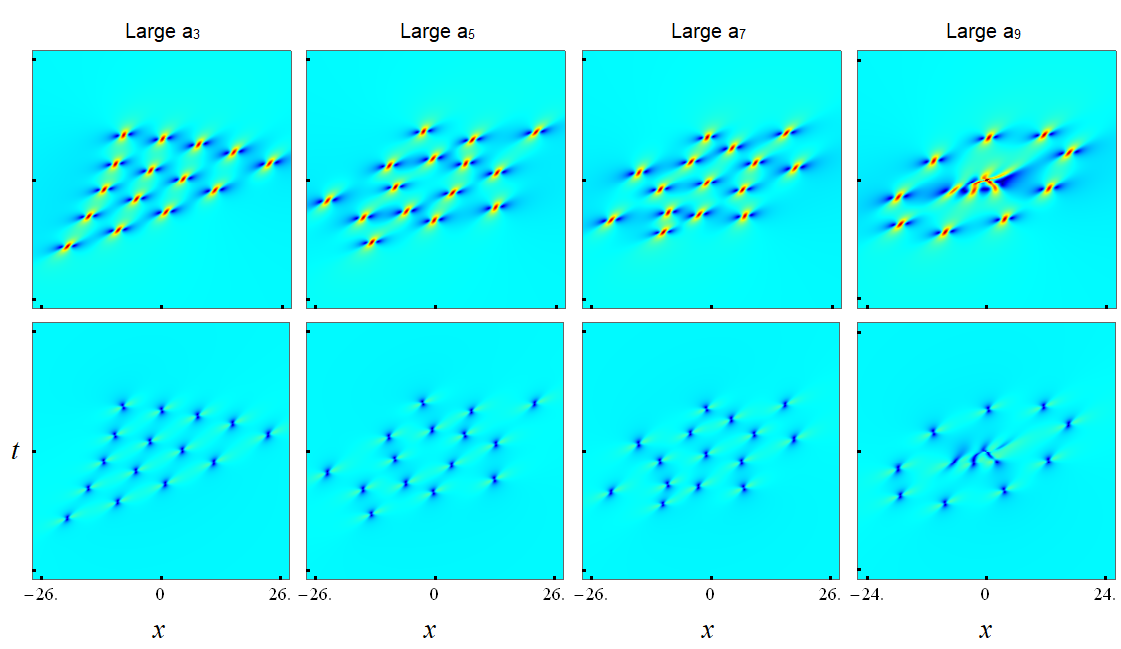}
\caption{True fifth-order Manakov rogue patterns. The system parameters, internal parameters and $(x,t)$ intervals for these true solutions are identical to those in the theoretically predicted patterns shown in Fig.~\ref{f:Mana1}.  }  \label{f:Mana2}
\end{center}
\end{figure}

Now, we compare these predicted fifth-order Manakov rogue patterns to the true solutions. Under identical system and internal parameter choices and $(x,t)$ intervals as those in Fig.~\ref{f:Mana1}, the true Manakov rogue patterns are displayed in Fig.~\ref{f:Mana2}. It is seen that the true rogue patterns closely mimic the predicted ones on all major aspects such as the overall shapes, orientations, and center-rogue-wave profiles. Some minor differences do exist, such as the three sides of the triangular true-rogue patterns in the first column of Fig.~\ref{f:Mana2} are a little more curvy than the predicted ones in the first column of Fig.~\ref{f:Mana1}. But those differences will diminish if the large parameter $a_3$ in those panels gets larger. Quantitatively, we have also verified the order of accuracy of our analytical predictions as stated in Theorem~3, with details omitted.

\section{Proofs of the three theorems} \label{sec:proof}
Now, we prove our main theorems on rogue-pattern predictions stated in Sec.~\ref{s:prediction} for the three underlying integrable systems when one of the internal parameters in their rogue waves is large.

\subsection{Proof of Theorem 1 for the GDNLS equations}

The proof of Theorem 1 for the GDNLS equations is quite similar to that we developed in Ref.~\cite{YangNLS2021} for the NLS equation, since the bilinear GDNLS rogue waves in Lemma~1 have similar structures as those for the NLS equation. Thus, our proof here will be brief.

When $|a_{2m+1}|\gg 1$ and the other parameters $O(1)$ in the GDNLS rogue wave solution (\ref{BilinearTrans2}),
at $(x,t)$ where $\sqrt{x^2+t^2}=O\left(|a_{2m+1}|^{1/(2m+1)}\right)$, we have \cite{YangNLS2021}
\begin{eqnarray}\label{Skasym1}
S_{j}(\textbf{\emph{x}}^{+}(n,k) +\nu \textbf{\emph{s}})
= S_j(\textbf{v}) \left[1+O\left(a_{2m+1}^{-1/(2m+1)}\right)\right], \qquad  |a_{2m+1}| \gg 1,
\end{eqnarray}
where
\[  \label{vdef}
\textbf{v}=\left(\sqrt{\alpha}x+\sqrt{\alpha}\left[(\alpha-1)+ {\rm i}\sqrt{\alpha}\right]t, \hspace{0.06cm} 0, \cdots, 0, a_{2m+1}, 0, \cdots\right).
\]
Notice that the first element in this $\textbf{v}$ vector is simply the leading-order terms of $x_1^+$ in Lemma 1 when $(x,t)$ are large, and the retention of only these leading-order terms in $x_1^+$ is sufficient here since we are only looking for the leading-order asymptotics in Eq.~(\ref{Skasym1}) at the moment. From the definition of Schur polynomials (\ref{Schurdef}), we see that $S_j(\textbf{v})$ is related to the polynomial $p_{j}^{[m]}(z)$ in Eq.~(\ref{pkmz}) as
\begin{equation} \label{Skorder}
S_j(\textbf{v})=\Omega^{j}p_{j}^{[m]}(z),
\end{equation}
where $\Omega$ was defined in Eq.~(\ref{defOmega}), and
\begin{equation} \label{Az}
z=\Omega^{-1}\left(\sqrt{\alpha}x+\sqrt{\alpha}\left[(\alpha-1)+ {\rm i}\sqrt{\alpha}\right]t\right).
\end{equation}
Similar relations can also be obtained for $S_{j}(\textbf{\emph{x}}^{-}(n,k) +\nu \textbf{\emph{s}})$. Using these formulae and following identical steps as in Ref.~\cite{YangNLS2021}, we find that
\begin{equation} \label{sigmanmax}
\sigma_{n,k} \sim  \chi_1 \hspace{0.01cm} \hspace{0.05cm} |a_{2m+1}|^{\frac{N(N+1)}{2m+1}} \left|Q_{N}^{[m]}(z)\right|^2,
\quad \quad |a_{2m+1}| \gg 1,
\end{equation}
where
\begin{equation*}
\chi_1=c_N^{-2} \hspace{0.05cm} \left(\frac{1}{2}\right)^{N(N-1)} \left(\frac{2m+1}{2^{2m}}\right)^{\frac{N(N+1)}{2m+1}}.
\end{equation*}
Since $\chi_1$ is independent of $n$ and $k$, the above equation shows that for large $a_{2m+1}$, $\sigma_{-1,1}/\sigma_{0,0}\sim 1$, i.e., the solution $|u_N(x,t)|$ in Eq.~(\ref{BilinearTrans2}) is on the unit background, except at or near $(x, t)$ locations $\left(\hat{x}_{0}, \hat{t}_{0}\right)$ where
\[  \label{z0def}
z_0=\Omega^{-1}\left(\sqrt{\alpha}\hat{x}_0+\sqrt{\alpha}\left[(\alpha-1)+ {\rm i}\sqrt{\alpha}\right]\hat{t}_0 \right)
\]
is a root of the polynomial $Q_{N}^{[m]}(z)$. These $\left(\hat{x}_{0}, \hat{t}_{0}\right)$ locations can be derived explicitly from the the above equation, and their expressions are as given in Eqs.~(\ref{x0Th1})-(\ref{t0Th1}) of Theorem~1.

Next, we show that when $(x, t)$ is in the neighborhood of each of the $\left(\hat{x}_{0}, \hat{t}_{0}\right)$ locations given by Eqs.~(\ref{x0Th1})-(\ref{t0Th1}), i.e., when $\left[(x-\hat{x}_{0})^2+(t-\hat{t}_{0})^2\right]^{1/2}=O(1)$, the rogue wave $u_N(x,t)$ in Eq. (\ref{BilinearTrans2}) approaches a fundamental rogue wave $\hat{u}_1(x-\hat{x}_{0}, t-\hat{t}_{0})\hspace{0.06cm} e^{{\rm i}(1-\gamma-\alpha) x-{\frac{\rm i}{2}}\left[\alpha^2+2(\gamma-2)\alpha+1-\gamma\right]t}$ for large $a_{2m+1}$, where the function $|\hat{u}_1(x,t)|$ is given in Eq.~(\ref{u1hatGDNLS}). For this purpose, we notice that when $(x, t)$ is in the neighborhood of $\left(\hat{x}_{0}, \hat{t}_{0}\right)$, we have a more refined asymptotics for $S_j(\textbf{\emph{x}}^{+}(n,k)+\nu \textbf{\emph{s}})$ as \cite{YangNLS2021}
\[ \label{Skxnasym2}
S_j(\textbf{\emph{x}}^{+}(n,k) +\nu \textbf{\emph{s}}) = S_j(\hat{\textbf{v}}) \left[1+O\left(a_{2m+1}^{-2/(2m+1)}\right)\right],
\]
where
\[\label{vhatdef}
\hat{\textbf{v}}=(x_1^+, 0, \cdots, 0, a_{2m+1}, 0, \cdots),
\]
and $x_1^+$ is as given in Lemma 1. Notice that this new asymptotics boasts a relative error that is an order smaller than that of the previous asymptotics (\ref{Skasym1}); thus it is more accurate. This more accurate asymptotics is necessary because the previous asymptotics (\ref{Skasym1}) gives only zero contribution to $\sigma_{n,k}$ in the neighborhood of $\left(\hat{x}_{0}, \hat{t}_{0}\right)$, as we have shown in Eq.~(\ref{sigmanmax}) above. To arrive at this more refined asymptotics (\ref{Skxnasym2}), we have utilized the fact of $s_1=0$ and $x_2^+=0$ in our bilinear rogue wave solution of Lemma 1. In addition, we have kept all terms of $x_1^+$ in $\hat{\textbf{v}}$ now, not just its leading-order terms.

The polynomial $S_j(\hat{\textbf{v}})$ in Eq.~(\ref{Skxnasym2}) is related to $p_{j}^{[m]}(z)$ of Eq.~(\ref{pkmz}) as $S_j(\hat{\textbf{v}})=\Omega^{j}p_{j}^{[m]}\left(\Omega^{-1}x_1^+\right)$, where $\Omega$ is given in Eq.~(\ref{defOmega}). Inserting this relation into (\ref{Skxnasym2}), we get a refined asymptotics for $S_j(\textbf{\emph{x}}^{+}(n,k) +\nu \textbf{\emph{s}})$ through polynomials $p_{j}^{[m]}(z)$. Similar refined asymptotics can also be obtained for $S_j(\textbf{\emph{x}}^{-}(n,k) +\nu \textbf{\emph{s}})$.

Using these refined asymptotics of $S_j(\textbf{\emph{x}}^{\pm}(n,k) +\nu \textbf{\emph{s}})$ and following the same steps as in Ref.~\cite{YangNLS2021}, we find that
\[ \label{sigmanxt5}
\sigma_{n,k}(x,t) = \hat{\chi}_1 \hspace{0cm} \left|\left[Q_{N}^{[m]}\right]'\hspace{-0.08cm}(z_0)\right|^2 |a_{2m+1}|^{\frac{N(N+1)-2}{(2m+1)}}
\left[ x_1^+(x-\hat{x}_{0}, t-\hat{t}_{0}; n,k ) x_1^-(x-\hat{x}_{0}, t-\hat{t}_{0}; n,k ) +\frac{1}{4} \right]\left[1+O\left(a_{2m+1}^{-1/(2m+1)}\right)\right],
\]
where $\hat{\chi}_1= [(2m+1)2^{-2m}]^{-2/(2m+1)}\chi_1$. Finally, we recall our assumption on the simplicity of the nonzero roots in Yablonskii-Vorob'ev polynomials in Sec.~\ref{sec:YV}, which implies that
$\left[Q_{N}^{[m]}\right]'\hspace{-0.08cm}(z_0)\ne 0$. This indicates that the above leading-order asymptotics for $\sigma_{n,k}(x,t)$ does not vanish. Therefore, when $a_{2m+1}$ is large and $(x, t)$ in the neighborhood of $\left(\hat{x}_{0}, \hat{t}_{0}\right)$, we insert the above equation into the rogue-wave solution (\ref{BilinearTrans2}) of Lemma 1 and get
\begin{eqnarray}
&& u_N(x,t) = e^{{\rm i}(1-\gamma-\alpha) x -{\frac{\rm i}{2}}\left[\alpha^2+2(\gamma-2)\alpha+1-\gamma\right]t}\hspace{0.05cm}\frac{(\sigma_{0,0}^*)^{\gamma-1}\sigma_{-1,1}}
{\sigma_{0,0}^\gamma}   \nonumber \\
&& =\hat{u}_1(x-\hat{x}_{0}, t-\hat{t}_{0})\hspace{0.06cm} e^{{\rm i}(1-\gamma-\alpha) x-{\frac{\rm i}{2}}\left[\alpha^2+2(\gamma-2)\alpha+1-\gamma\right]t}+ O\left(a_{2m+1}^{-1/(2m+1)}\right),
\end{eqnarray}
where
\[
\hat{u}_1(x, t)=\frac{
\left[ x_1^+(x, t; 0,0 ) x_1^-(x, t; 0,0 )+\frac{1}{4}\right]^{*(\gamma-1)}
\left[ x_1^+(x, t; -1,1 ) x_1^-(x, t; -1,1 )+\frac{1}{4}\right]}
{\left[ x_1^+(x, t; 0,0 ) x_1^-(x, t; 0,0 )+\frac{1}{4}\right]^{\gamma}}
\]
is the fundamental GDNLS rogue wave, whose amplitude function $|\hat{u}_1(x, t)|$ can be simplified as Eq.~(\ref{u1hatGDNLS}), and the error of this prediction is $O\left(a_{2m+1}^{-1/(2m+1)}\right)$.

Regarding the proof for the GDNLS rogue pattern near the origin in Theorem~1, it is identical to that for the NLS equation in Ref.~\cite{YangNLS2021}. Theorem~1 is then proved.

\subsection{Proof of Theorem 2 for the Boussinesq equation}

The proof for Theorem 2 is along similar lines of the previous proof but with small modifications.

When $|a_{2m+1}|\gg 1$ and the other parameters $O(1)$ in the Boussinesq rogue wave solution (\ref{Boussi_rogue}), at $(x,t)$ where $\sqrt{x^2+t^2}=O\left(|a_{2m+1}|^{1/(2m+1)}\right)$, we have
\begin{eqnarray}\label{Skasym1B}
S_j(\textbf{\emph{x}}^{+}  +\nu \textbf{\emph{s}}) = S_j(\textbf{v}) \left[1+O\left(a_{2m+1}^{-1/(2m+1)}\right)\right], \qquad  |a_{2m+1}| \gg 1,
\end{eqnarray}
where
\[  \label{vdefB}
\textbf{v}=\left(x_1^+, \hspace{0.06cm} 0, \cdots, 0, a_{2m+1}, 0, \cdots\right),
\]
and $x_1^+(x,t)=(\hspace{-0.04cm}\sqrt{3}\hspace{0.04cm}{\rm{i}}/6) \hspace{0.04cm} (x+{\rm{i}}t)$ from Lemma 2. Unlike the earlier GDNLS case, all terms in $x_1^+$ here are of the same order for large $(x,t)$; hence we keep all of them. But we did drop the $\nu s_1$ term of $x_1^{+}  +\nu s_1$ in $\textbf{v}$ even though $s_1\ne 0$ now [see Eq.~(\ref{skB})], because that $\nu s_1$ term is much smaller than $x_1^+$ and thus does not contribute to the leading order. The Schur polynomial $S_j(\textbf{v})$ is related to the polynomial $p_{j}^{[m]}(z)$ of Eq.~(\ref{pkmz}) in the same way as Eq.~(\ref{Skorder}), except that $z$ is now replaced by $z=\Omega^{-1}x_1^+$, where $\Omega$ is as given in Eq.~(\ref{defOmega}). Thus,
\[
S_j(\textbf{\emph{x}}^{+}  +\nu \textbf{\emph{s}}) \sim \Omega^{j}p_{j}^{[m]}(z), \qquad  |a_{2m+1}| \gg 1.
\]
Similarly, using the definition of Schur polynomials (\ref{Schurdef}) and the fact of $x_1^{-}=-(x_1^{+})^*$ from Lemma 2, we get
\[
S_j(\textbf{\emph{x}}^{-}  +\nu \textbf{\emph{s}}) \sim (-\Omega^*)^{j} p_{j}^{[m]}\left(z^*\right), \qquad  |a_{2m+1}| \gg 1.
\]
Using these formulae and following the same steps as in Ref.~\cite{YangNLS2021}, we find that $\sigma$ of Eq.~(\ref{SigmanAlg}) has the asymptotics
\begin{equation} \label{sigmanmaxB}
\sigma \sim \chi_2 \hspace{0.03cm} |a_{2m+1}|^{\frac{N(N+1)}{2m+1}} \left|Q_{N}^{[m]}(z)\right|^2,
\quad \quad |a_{2m+1}| \gg 1,
\end{equation}
where $\chi_2$ is a $(N,m)$-dependent constant. This asymptotics shows that the solution $u_N(x,t)$ in Eq.~(\ref{Boussi_rogue}) would be approximately zero, except at or near $(x, t)$ locations $\left(\tilde{x}_{0}, \tilde{t}_{0}\right)$ where $\Omega^{-1} x_1^+$ is a root of the polynomial $Q_{N}^{[m]}(z)$, i.e.,
\[  \label{z0defB}
z_0=\Omega^{-1}  \frac{\sqrt{3}\hspace{0.04cm}\rm{i}}{6}(\tilde{x}_{0}+{\rm{i}}\hspace{0.03cm} \tilde{t}_{0})
\]
is a root of the polynomial $Q_{N}^{[m]}(z)$. These $\left(\tilde{x}_{0}, \tilde{t}_{0}\right)$ locations can be derived explicitly from the the above equation as
\begin{eqnarray}
&& \tilde{x}_{0}=\Re\left[ z_{0} \left(-2\textrm{i} \sqrt{3}\right)\Omega \right],
\quad \tilde{t}_{0}=\Im\left[z_{0} \left(-2\textrm{i} \sqrt{3}\right) \Omega \right].
\end{eqnarray}
Notice that these $\left(\tilde{x}_{0}, \tilde{t}_{0}\right)$ values match the leading-order terms of $\left(\hat{x}_{0}, \hat{t}_{0}\right)$ given in Theorem~2.

Next, we show that when $(x, t)$ is in the neighborhood of each of the $\left(\tilde{x}_{0}, \tilde{t}_{0}\right)$ locations given by the above equation, i.e., when $\left[(x-\tilde{x}_{0})^2+(t-\tilde{t}_{0})^2\right]^{1/2}=O(1)$, the Boussinesq rogue wave $u_N(x,t)$ in Eq.~(\ref{Boussi_rogue}) under large $a_{2m+1}$ approaches a fundamental rogue wave, whose location is not exactly at $\left(\tilde{x}_{0}, \tilde{t}_{0}\right)$ but is close to it. For this purpose, we use a refined asymptotics (\ref{Skxnasym2}) for $S_j(\textbf{\emph{x}}^{+}+\nu \textbf{\emph{s}})$ when $(x, t)$ is in the neighborhood of $\left(\tilde{x}_{0}, \tilde{t}_{0}\right)$, where the vector $\hat{\textbf{v}}$ is now
\[\label{vhatdefB}
\hat{\textbf{v}}=(x_1^{+} + \nu s_1, 0, \cdots, 0, a_{2m+1}, 0, \cdots).
\]
Compared to its GDNLS counterpart (\ref{vhatdef}) and a similar one for the NLS equation in Ref.~\cite{YangNLS2021}, the vector $\hat{\textbf{v}}$ here contains a new term $\nu s_1$. This term was absent in the GDNLS and NLS cases because $s_1$ vanishes there. This new $\nu s_1$ term in $\hat{\textbf{v}}$ will introduce an additional $O(1)$ contribution to the fundamental-rogue-wave location $\left(\hat{x}_{0}, \hat{t}_{0}\right)$ in Boussinesq rogue patterns, as we will see below.

The polynomial $S_j(\hat{\textbf{v}})$ is related to $p_{j}^{[m]}(z)$ of Eq.~(\ref{pkmz}). Using that relation, we can rewrite the asymptotics of $S_j(\textbf{\emph{x}}^{+}+\nu \textbf{\emph{s}})$ as
\begin{equation} \label{Skorder2B}
S_j(\textbf{\emph{x}}^{+}+\nu \textbf{\emph{s}}) = \Omega^{j}p_{j}^{[m]}\left(z+ \nu s_1 \Omega^{-1}\right) \left[1+O\left(a_{2m+1}^{-2/(2m+1)}\right)\right],
\end{equation}
where $\Omega$ is given in Eq.~(\ref{defOmega}), and $z=\Omega^{-1}x_1^+$ as mentioned before. Similarly, we have \begin{equation} \label{Skorder2B2}
S_j(\textbf{\emph{x}}^{-}+\nu \textbf{\emph{s}}) = (-\Omega^*)^{j} \left[p_{j}^{[m]}\left(z+ \nu s_1 \Omega^{-1}\right)\right]^*\left[1+O\left(a_{2m+1}^{-2/(2m+1)}\right)\right]
\end{equation}
in view that $s_1$ is purely imaginary [see Eq.~(\ref{skB})].

Using these refined asymptotics of $S_j(\textbf{\emph{x}}^{\pm} +\nu \textbf{\emph{s}})$, we can now estimate the leading order terms of $\sigma(x,t)$ in Eq.~(\ref{SigmanAlg}) when $(x, t)$ is in the neighborhood of $\left(\tilde{x}_{0}, \tilde{t}_{0}\right)$. For this purpose, we use determinant identities and the Laplace expansion to rewrite $\sigma(x,t)$ as \cite{OhtaJY2012}
\begin{eqnarray} \label{sigmanLapB}
&& \hspace{-0.8cm} \sigma(x,t)=\sum_{0\leq\nu_{1} < \nu_{2} < \cdots < \nu_{N}\leq 2N-1}
\det_{1 \leq i, j\leq N} \left[\left(\frac{\sqrt{3}\hspace{0.04cm}\rm{i}}{6}\right)^{\nu_j} S_{2i-1-\nu_j}(\textbf{\emph{x}}^{+} +\nu_j \textbf{\emph{s}}) \right]  \times \det_{1 \leq i, j\leq N}\left[\left(\frac{\sqrt{3}\hspace{0.04cm}\rm{i}}{6}\right)^{\nu_j} S_{2i-1-\nu_j}(\textbf{\emph{x}}^{-} + \nu_j \textbf{\emph{s}})\right] \nonumber,\\
&&
\end{eqnarray}
where $S_j\equiv 0$ when $j<0$. When $a_{2m+1}$ is large, $\Omega$ is large. In this case, contributions to leading-order terms in $\sigma(x,t)$ come from two index choices, one being $\nu=(0, 1, \cdots, N-1)$, and the other being $\nu=(0, 1, \cdots, N-2, N)$.

For the first index choice of $\nu=(0, 1, \cdots, N-1)$, i.e., $\nu_j=j-1$, we have
\[ \label{detasym1B}
\hspace{-0.5cm}\det_{1 \leq i, j\leq N} \left[\left(\frac{\sqrt{3}\hspace{0.04cm}\rm{i}}{6}\right)^{\nu_j} S_{2i-1-\nu_j}(\textbf{\emph{x}}^{+} +\nu_j \textbf{\emph{s}}) \right] = \left(\frac{\sqrt{3}\hspace{0.04cm}\rm{i}}{6}\right)^{\frac{N(N-1)}{2}}\Omega^{\frac{N(N+1)}{2}}
\det_{1 \leq i, j\leq N} \left[p_{2i-j}^{[m]}\left(z+ (j-1) s_1 \Omega^{-1}\right) \right] \left[1+O\left(a_{2m+1}^{-2/(2m+1)}\right)\right].
\]
Since $z=\Omega^{-1}x_1^+$, which is a linear function of $x$ and $t$, and $z_0=\Omega^{-1}x_1^+(\tilde{x}_{0}, \tilde{t}_{0})$, we can rewrite $z+ (j-1) s_1 \Omega^{-1}$ as
\[
z+ (j-1) s_1 \Omega^{-1} = z_0 + \left[x_1^+(x-\tilde{x}_{0}, t-\tilde{t}_{0})+(j-1) s_1\right]\Omega^{-1} .
\]
Then, expanding $p_{2i-j}^{[m]}\left(z+ (j-1) s_1 \Omega^{-1}\right)$ around $z_0$, we get
\[
p_{2i-j}^{[m]}\left(z+ (j-1) s_1 \Omega^{-1}\right)=p_{2i-j}^{[m]}(z_0)+\left(p_{2i-j}^{[m]}\right)'
\hspace{-0.08cm}(z_0)\left[
x_1^+(x-\tilde{x}_{0}, t-\tilde{t}_{0})+(j-1)s_1\right]\Omega^{-1}+O(\Omega^{-2}).
\]
Substituting this expansion into the determinant of the above equation (\ref{detasym1B}), the $O(1)$ term of this determinant is $Q_{N}^{[m]}(z_0)$, which is zero since $z_0$ is a root of this Yablonskii-Vorob'ev polynomial. To derive the remaining terms, we recall the property (\ref{pkpkp1}) of $p_{j}^{[m]}(z)$ polynomials, which yields $\left(p_{2i-j}^{[m]}\right)'\hspace{-0.08cm}(z_0)=p_{2i-j-1}^{[m]}(z_0)$. Utilizing this relation, we find that many terms in the expansion of the determinant in (\ref{detasym1B}) vanish, and the only dominant contribution is
\begin{eqnarray*}
&&\det_{1 \leq i, j\leq N} \left[p_{2i-j}^{[m]}\left(z+ (j-1) s_1 \Omega^{-1}\right) \right]  \\
&& =\det_{1 \leq i\leq N} \left[p_{2i-1}^{[m]}(z_0), p_{2i-2}^{[m]}(z_0), \cdots, p_{2i-N+1}^{[m]}(z_0), p_{2i-N-1}^{[m]}(z_0)\left[x_1^+(x-\tilde{x}_{0}, t-\tilde{t}_{0})+(N-1)s_1\right]\right]\Omega^{-1}+O(\Omega^{-2}) \\
&& =c_N^{-1}\left[Q_{N}^{[m]}\right]'\hspace{-0.08cm}(z_0) \left[x_1^+(x-\tilde{x}_{0}, t-\tilde{t}_{0})+(N-1)s_1\right] \hspace{0.08cm} \Omega^{-1}+O(\Omega^{-2}),
\end{eqnarray*}
where $c_N$ is given below Eq.~(\ref{GeneralYablonski}). In the last step, the fact of the determinant (\ref{GeneralYablonski}) of $Q_{N}^{[m]}(z)$ being Wronskian has been used. Recalling that the $s_1$ value from Eq.~(\ref{skB}) is purely imaginary, we can absorb the $(N-1)s_1$ term of the above equation into its $\tilde{x}_{0}$ term and get
\[
\det_{1 \leq i, j\leq N} \left[p_{2i-j}^{[m]}\left(z+ (j-1) s_1 \Omega^{-1}\right) \right]
=c_N^{-1}\left[Q_{N}^{[m]}\right]'\hspace{-0.08cm}(z_0) \hspace{0.06cm} x_1^+(x-\hat{x}_{0}, t-\hat{t}_{0})  \hspace{0.1cm} \Omega^{-1}+O(\Omega^{-2}),
\]
where $\left(\hat{x}_{0}, \hat{t}_{0}\right)$ are as given in Theorem~2. Inserting this result into (\ref{detasym1B}) and noticing that $\Omega^{-1}=O(a_{2m+1}^{-1/(2m+1)})$, we get
\[
\det_{1 \leq i, j\leq N} \left[\left(\frac{\sqrt{3}\hspace{0.04cm}\rm{i}}{6}\right)^{\nu_j} S_{2i-1-\nu_j}(\textbf{\emph{x}}^{+} +\nu_j \textbf{\emph{s}}) \right] = c_N^{-1} \left[Q_{N}^{[m]}\right]'\hspace{-0.08cm}(z_0) \left(\frac{\sqrt{3}\hspace{0.04cm}\rm{i}}{6}\right)^{\frac{N(N-1)}{2}}\Omega^{\frac{N(N+1)-2}{2}}
x_1^+(x-\hat{x}_{0}, t-\hat{t}_{0}) \left[1+O\left(a_{2m+1}^{-1/(2m+1)}\right)\right].
\]
Similarly, we get
\[
\hspace{-0.75cm} \det_{1 \leq i, j\leq N} \left[\left(\frac{\sqrt{3}\hspace{0.04cm}\rm{i}}{6}\right)^{\nu_j} S_{2i-1-\nu_j}(\textbf{\emph{x}}^{-} +\nu_j \textbf{\emph{s}}) \right] = c_N^{-1} \left[Q_{N}^{[m]}\right]'\hspace{-0.08cm}(z_0^*) \left(\frac{\sqrt{3}\hspace{0.04cm}\rm{i}}{6}\right)^{\frac{N(N-1)}{2}}\left(-\Omega^*\right)^{\frac{N(N+1)-2}{2}}
x_1^{-}(x-\hat{x}_{0}, t-\hat{t}_{0}) \left[1+O\left(a_{2m+1}^{-1/(2m+1)}\right)\right].
\]
Then, the contribution to Eq.~(\ref{sigmanLapB}) from the first index choice of $\nu=(0, 1, \cdots, N-1)$ is
\[
\hat{\chi}_2 \hspace{0.04cm} \left|\left[Q_{N}^{[m]}\right]'\hspace{-0.08cm}(z_0) \right|^2 |a_{2m+1}|^{\frac{N(N+1)-2}{2m+1}}
x_1^{+}(x-\hat{x}_{0}, t-\hat{t}_{0}) \hspace{0.06cm} x_1^{-}(x-\hat{x}_{0}, t-\hat{t}_{0})  \left[1+O\left(a_{2m+1}^{-1/(2m+1)}\right)\right],
\]
where
\[ \label{chi2hat}
\hat{\chi}_2= c_N^{-2} (-1)^{N^2-1}\left(\frac{1}{12}\right)^{\frac{N(N-1)}{2}} \left(\frac{2m+1}{2^{2m}}\right)^{\frac{N(N+1)-2}{2m+1}}.
\]

Next, we calculate the contribution to $\sigma(x,t)$ in Eq.~(\ref{sigmanLapB}) from the second index choice of $\nu=(0, 1, \cdots, N-2, N)$. In this case, utilizing Eq.~(\ref{Skorder2B}), we get
\[ \label{detasym1B2}
\hspace{-0.6cm}\det_{1 \leq i, j\leq N} \left[\left(\frac{\sqrt{3}\hspace{0.04cm}\rm{i}}{6}\right)^{\nu_j} S_{2i-1-\nu_j}(\textbf{\emph{x}}^{+} +\nu_j \textbf{\emph{s}}) \right] = \left(\frac{\sqrt{3}\hspace{0.04cm}\rm{i}}{6}\right)^{\frac{N(N-1)+2}{2}}\Omega^{\frac{N(N+1)-2}{2}}
\det_{1 \leq i, j\leq N} \left[p_{2i-1-\nu_j}^{[m]}\left(z+ \nu_j s_1 \Omega^{-1}\right) \right] \left[1+O\left(a_{2m+1}^{-2/(2m+1)}\right)\right].
\]
The leading-order asymptotics of this determinant can be directly obtained by neglecting the $\nu_j s_1 \Omega^{-1}$ terms and replacing $z$ by $z_0$, which gives
\begin{eqnarray}
&& \det_{1 \leq i, j\leq N} \left[\left(\frac{\sqrt{3}\hspace{0.04cm}\rm{i}}{6}\right)^{\nu_j} S_{2i-1-\nu_j}(\textbf{\emph{x}}^{+} +\nu_j \textbf{\emph{s}}) \right] = \left(\frac{\sqrt{3}\hspace{0.04cm}\rm{i}}{6}\right)^{\frac{N(N-1)+2}{2}}\Omega^{\frac{N(N+1)-2}{2}}
\det_{1 \leq i, j\leq N} \left[p_{2i-1-\nu_j}^{[m]}\left(z_0\right) \right] \left[1+O\left(a_{2m+1}^{-1/(2m+1)}\right)\right]   \nonumber \\
&& \hspace{6.3cm} = \left(\frac{\sqrt{3}\hspace{0.04cm}\rm{i}}{6}\right)^{\frac{N(N-1)+2}{2}}\Omega^{\frac{N(N+1)-2}{2}}
\left[Q_{N}^{[m]}\right]'\hspace{-0.08cm}(z_0) \left[1+O\left(a_{2m+1}^{-1/(2m+1)}\right)\right].  \label{detasym1B3}
\end{eqnarray}
Similarly,
\begin{eqnarray}
&& \hspace{-1.0cm}\det_{1 \leq i, j\leq N} \left[\left(\frac{\sqrt{3}\hspace{0.04cm}\rm{i}}{6}\right)^{\nu_j} S_{2i-1-\nu_j}(\textbf{\emph{x}}^{-} +\nu_j \textbf{\emph{s}}) \right] = \left(\frac{\sqrt{3}\hspace{0.04cm}\rm{i}}{6}\right)^{\frac{N(N-1)+2}{2}}(-\Omega^*)^{\frac{N(N+1)-2}{2}}
\left[Q_{N}^{[m]}\right]'\hspace{-0.08cm}(z_0^*) \left[1+O\left(a_{2m+1}^{-1/(2m+1)}\right)\right].  \label{detasym1B4}
\end{eqnarray}
Then, the contribution to Eq.~(\ref{sigmanLapB}) from the second index choice is
\[
\hat{\chi}_2 \hspace{0.04cm} \left|\left[Q_{N}^{[m]}\right]'\hspace{-0.08cm}(z_0) \right|^2 |a_{2m+1}|^{\frac{N(N+1)-2}{2m+1}}
\left[-\frac{1}{12}+O\left(a_{2m+1}^{-1/(2m+1)}\right)\right],
\]
where $\hat{\chi}_2$ is given in Eq.~(\ref{chi2hat}). Combining both contributions to $\sigma(x,t)$ of Eq.~(\ref{sigmanLapB}), we get
\begin{eqnarray}
&& \sigma(x,t) = \hat{\chi}_2 \hspace{0.04cm} \left|\left[Q_{N}^{[m]}\right]'\hspace{-0.08cm}(z_0) \right|^2 |a_{2m+1}|^{\frac{N(N+1)-2}{2m+1}}
\left[x_1^{+}(x-\hat{x}_{0}, t-\hat{t}_{0}) \hspace{0.06cm} x_1^{-}(x-\hat{x}_{0}, t-\hat{t}_{0}) -\frac{1}{12}\right]  \left[1+O\left(a_{2m+1}^{-1/(2m+1)}\right)\right] \nonumber \\
&&\hspace{1cm}= -\frac{1}{12}\hat{\chi}_2 \hspace{0.04cm} \left|\left[Q_{N}^{[m]}\right]'\hspace{-0.08cm}(z_0) \right|^2 |a_{2m+1}|^{\frac{N(N+1)-2}{2m+1}}\left[(x-\hat{x}_{0})^2+(t-\hat{t}_{0})^2+1\right]  \left[1+O\left(a_{2m+1}^{-1/(2m+1)}\right)\right].
\end{eqnarray}
Recalling our assumption on the simplicity of the nonzero roots in Yablonskii-Vorob'ev polynomials in Sec.~\ref{sec:YV}, the above leading-order asymptotics for $\sigma(x,t)$ does not vanish. Therefore, when $a_{2m+1}$ is large and $(x, t)$ in the neighborhood of $\left(\hat{x}_{0}, \hat{t}_{0}\right)$, the above asymptotics and Eq.~(\ref{Boussi_rogue}) show that
\begin{eqnarray}
&& u_{N}(x,t)=2 \partial_{x}^2 \ln \sigma = 2 \partial_{x}^2 \ln \left[(x-\hat{x}_{0})^2+(t-\hat{t}_{0})^2+1\right]+ O\left(a_{2m+1}^{-1/(2m+1)}\right),
\end{eqnarray}
which proves Eq.~(\ref{Th2uN}) in Theorem 2.

Next, we prove Eq.~(\ref{ucenter2}) in Theorem 2 regarding the asymptotics of the rogue wave $u_N(x,t)$ in the neighborhood of the origin. In this case, we first rewrite the $N\times N$ determinant of $\sigma(x,t)$ in Eq.~(\ref{SigmanAlg}) as a $3N\times 3N$ determinant \cite{OhtaJY2012}
\[ \label{3Nby3Ndet2}
\sigma(x,t)=\left|\begin{array}{cc}
\textbf{O}_{N\times N} & \Phi_{N\times 2N} \\
-\Psi_{2N\times N} & \textbf{I}_{2N\times 2N} \end{array}\right|,
\]
where
\begin{eqnarray} \label{PhiPsihatB0}
&& \Phi_{i,j}=
\left(\frac{\sqrt{3}\hspace{0.04cm}\rm{i}}{6}\right)^{j-1}
S_{2i-j}\left[\textbf{\emph{x}}^{+}  + (j-1) \textbf{\emph{s}}\right], \quad \Psi_{i,j}=  \left(\frac{\sqrt{3}\hspace{0.04cm}\rm{i}}{6}\right)^{i-1} S_{2j-i}\left[\textbf{\emph{x}}^{-}  + (i-1) \textbf{\emph{s}}\right],
\end{eqnarray}
and $S_j\equiv 0$ when $j<0$. Then, we use relations similar to Eq. (58) of Ref.~\cite{YangNLS2021} to express matrix elements of $\Phi$ and $\Psi$ as powers of $a_{2m+1}$ and $a_{2m+1}^*$ respectively. Next, we perform the same row and column operations of Ref.~\cite{YangNLS2021} to the $\Phi$ and $\Psi$ matrices so that certain high-power terms of $a_{2m+1}$ and $a_{2m+1}^*$ are eliminated. Afterwards, we keep only the highest power terms of $a_{2m+1}$ in each row or column of the remaining $\Phi$ and $\Psi$ matrices. With these manipulations, $\sigma(x,t)$ is asymptotically reduced to the same expression (64) of Ref.~\cite{YangNLS2021}, which is
\[ \label{3Nby3Ndet3}
\sigma(x,t)=\hat{\beta} \hspace{0.06cm}
\left|\begin{array}{cc}
\textbf{O}_{N_0\times N_0} & \widehat{\Phi}_{N_0\times 2N_0} \\
-\widehat{\Psi}_{2N_0\times N_0} & \textbf{I}_{2N_0\times 2N_0} \end{array}\right| \left[1+O\left(a_{2m+1}^{-1}\right)\right],
\]
where $\hat{\beta}$ is a $(m, N)$-dependent constant multiplying a certain power of $|a_{2m+1}|$,
\begin{eqnarray} \label{PhiPsihatB}
&& \widehat{\Phi}_{i,j}=  \left(\frac{\sqrt{3}\hspace{0.04cm}\rm{i}}{6}\right)^{j-1}S_{2i-j}\left[\textbf{\emph{y}}^{+}  + (j-1+\nu_0) \textbf{\emph{s}}\right], \quad \widehat{\Psi}_{i,j}=  \left(\frac{\sqrt{3}\hspace{0.04cm}\rm{i}}{6}\right)^{i-1} S_{2j-i}\left[\textbf{\emph{y}}^{-}  + (i-1+\nu_0) \textbf{\emph{s}}\right],
\end{eqnarray}
vectors $\textbf{\emph{y}}^{\pm}$ are $\textbf{\emph{x}}^{\pm}$ without the $a_{2m+1}$ terms, i.e.,
\[
\textbf{\emph{y}}^{+}=\textbf{\emph{x}}^{+}-(0, \cdots, 0, a_{2m+1}, 0, \cdots), \quad
\textbf{\emph{y}}^{-}=\textbf{\emph{x}}^{-}+(0, \cdots, 0, a_{2m+1}^*, 0, \cdots),
\]
and $\nu_0=N-N_0$, where $N_0$ is a function of $(N,m)$ as explained in Sec.~\ref{sec:YV}.

For the NLS and GDNLS equations, $s_1=s_3=\cdots=s_{odd}=0$. Because of that, the determinant in Eq.~(\ref{3Nby3Ndet3}) could be further reduced to one with $\nu_0$ set as zero in Eq.~(\ref{PhiPsihatB}) due to the relation (65) in Ref.~\cite{YangNLS2021}. As a consequence, the asymptotics (\ref{3Nby3Ndet3}) would yield a lower-order rogue wave $u_{N_0}(x,t)$ in the neighborhood of the origin, whose internal parameters were identical to the corresponding ones in the original rogue wave $u_N(x,t)$. However, for the Boussinesq equation, $s_{odd}$ is not zero [see Eq.~(\ref{skB})]. Because of this, a slightly different treatment is needed. In this case, we split the vector $\textbf{\emph{s}}$ into two vectors of even and odd elements,
\[
\textbf{\emph{s}}=\textbf{\emph{s}}_{odd} + \textbf{\emph{s}}_{even},
\]
where $\textbf{\emph{s}}_{odd}=(s_1, 0, s_3, 0, \cdots)$, and $\textbf{\emph{s}}_{even}=(0, s_2, 0, s_4, \cdots)$. Then, using a relation similar to Eq.~(65) in Ref.~\cite{YangNLS2021}, we can show that the determinant in Eq.~(\ref{3Nby3Ndet3}) is equal to one whose matrix elements are reduced to
\begin{eqnarray} \label{PhiPsihatB2}
&& \hspace{-0.5cm}\widehat{\Phi}_{i,j}= \left(\frac{\sqrt{3}\hspace{0.04cm}\rm{i}}{6}\right)^{j-1} S_{2i-j}\left[\textbf{\emph{y}}^{+}  +
(j-1) \hspace{0.06cm} \textbf{\emph{s}}+\nu_0 \hspace{0.06cm} \textbf{\emph{s}}_{odd}\right], \quad \widehat{\Psi}_{i,j}=  \left(\frac{\sqrt{3}\hspace{0.04cm}\rm{i}}{6}\right)^{i-1} S_{2j-i}\left[\textbf{\emph{y}}^{-}  +(i-1)\hspace{0.06cm} \textbf{\emph{s}}+ \nu_0\hspace{0.06cm} \textbf{\emph{s}}_{odd}\right].
\end{eqnarray}
Lastly, we absorb $\nu_0\hspace{0.06cm}\textbf{\emph{s}}_{odd}$, i.e., $(N-N_0)\hspace{0.06cm}\textbf{\emph{s}}_{odd}$, into the vector $\textbf{\emph{y}}^{\pm}$ in the above equations by redefining parameters
\[\label{ParameterRela}
\hat{a}_{2r-1}=a_{2r-1}+\left(N-N_{0}\right)s_{2r-1}, \quad r=1, 2, \cdots, N_0.
\]
This absorbtion is possible since all $s_{odd}$ values are purely imaginary for the Boussinesq equation, see Note 2 below Lemma 2; hence absorbtions for $\textbf{\emph{y}}^{+}$ and $\textbf{\emph{y}}^{-}$ are consistent. The $\sigma(x,t)$ function (\ref{3Nby3Ndet3}) with such matrix elements (\ref{PhiPsihatB2}) then gives a $N_0$-th order rogue wave, whose internal parameters $\hat{a}_{2r-1}$ are related to ${a}_{2r-1}$ of the original $N$-th order rogue wave $u_N(x,t)$ through Eq.~(\ref{ParameterRela}), i.e., (\ref{akBoussi}) of Theorem 2, and the error of this lower-order rogue wave approximation is $O(|a_{2m+1}|^{-1})$ in view of Eq.~(\ref{3Nby3Ndet3}). Eq.~(\ref{ucenter2}) in Theorem 2 is then proved.

\subsection{Proof of Theorem 3 for the Manakov system}

The proof of Theorem 3 for the Manakov system has a lot in common with proofs for Theorems 1 and 2; so we will be brief here.

First, we consider the solution asymptotics away from the origin when $a_{2m+1}$ is large. In this case,
\begin{eqnarray}\label{Skasym1M}
&& S_{j}(\textbf{\emph{x}}^{+}(n,k) +\nu \textbf{\emph{s}})
= S_j(\textbf{v}) \left[1+O\left(a_{2m+1}^{-1/(2m+1)}\right)\right], \hspace{1cm}  |a_{2m+1}| \gg 1,  \\
&& S_{j}(\textbf{\emph{x}}^{-}(n,k) +\nu \textbf{\emph{s}}^*)
= S_j(\textbf{v}^*) \left[1+O\left(a_{2m+1}^{-1/(2m+1)}\right)\right],  \hspace{0.75cm}  |a_{2m+1}| \gg 1,
\end{eqnarray}
where
\[  \label{vdefM}
\textbf{v}=\left(p_{1} x + 2 \textrm{i} p_{0} p_{1} t, \hspace{0.06cm} 0, \cdots, 0, a_{2m+1}, 0, \cdots\right),
\]
whose first element is the leading-order terms of $x_1^+$ in Lemma 3. Using these asymptotics and repeating similar steps as before, we find that the solution $[|u_{1,N}(x,t)|, |u_{2,N}(x,t)|]$ would be on the uniform background except at or near $(x,t)$ locations $\left(\tilde{x}_{0}, \tilde{t}_{0}\right)$ where
\[  \label{z0defM}
z_0=\Omega^{-1}\left(p_{1} \tilde{x}_{0}+ 2 \textrm{i} p_{0} p_{1} \tilde{t}_{0} \right)
\]
is a root of the polynomial $Q_{N}^{[m]}(z)$. This $\left(\tilde{x}_{0}, \tilde{t}_{0}\right)$ is the leading-order approximate location of nontrivial rogue dynamics, and they can be solved explicitly as
\begin{eqnarray*}
\tilde{x}_{0}=\frac{1}{\Re(p_0)}\Re\left[ \frac{z_{0}\Omega}{p_{1}}p_{0}^* \right],  \quad
\tilde{t}_{0}=\frac{1}{2 \Re(p_0)} \Im\left[ \frac{z_{0}\Omega}{p_{1}} \right].
\end{eqnarray*}

Next, we use a refined asymptotics to show that near the above location is a fundamental Manakov rogue wave, whose position is the above $\left(\tilde{x}_{0}, \tilde{t}_{0}\right)$ plus a certain $O(1)$ shift. We start with the refined asymptotics
\begin{eqnarray} \label{Skorder2M}
S_j(\textbf{\emph{x}}^{+}(n,k)+\nu \textbf{\emph{s}}) = \Omega^{j}p_{j}^{[m]}\left[z^{+} + \nu s_1 \Omega^{-1}\right] \left[1+O\left(a_{2m+1}^{-2/(2m+1)}\right)\right],
\end{eqnarray}
where $z^{+}=\Omega^{-1}x_1^+$, and $x_1^{+}(x,t;n,k)$ is defined in Lemma 3. Performing similar calculations as in the proof of Theorem 2 for the Boussinesq equation, we have
\begin{eqnarray*}
&&\det_{1 \leq i, j\leq N} \left[p_{2i-j}^{[m]}\left(z^++ (j-1) s_1 \Omega^{-1}\right) \right]  =c_N^{-1}\left[Q_{N}^{[m]}\right]'\hspace{-0.08cm}(z_0) \left[
x_1^+(x-\tilde{x}_{0}, t-\tilde{t}_{0}; n,k )+(N-1)s_1\right]  \hspace{0.08cm} \Omega^{-1}+O(\Omega^{-2}).
\end{eqnarray*}
We can absorb the $(N-1)s_1$ term of the above equation into its $\tilde{x}_{0}$ and $\tilde{t}_{0}$ terms, so that it becomes
\begin{eqnarray} \label{pkmzOmegaM}
\det_{1 \leq i, j\leq N} \left[p_{2i-j}^{[m]}\left(z^{+} + (j-1) s_1 \Omega^{-1}\right) \right]  =c_N^{-1}\left[Q_{N}^{[m]}\right]'\hspace{-0.08cm}(z_0) \hspace{0.07cm}
x_1^+(x-\hat{x}_{0}, t-\hat{t}_{0}; n,k ) \hspace{0.1cm} \Omega^{-1}+O(\Omega^{-2}),
\end{eqnarray}
where $(\hat{x}_{0}, \hat{t}_{0})$ are as given in Eqs.~(\ref{x0Mana})-(\ref{t0Mana}) of Theorem 3. This $(\hat{x}_{0}, \hat{t}_{0})$ is the improved, and asymptotically more accurate, location of a fundamental rogue wave, which we will see in a moment.

The $\sigma_{n,k}$ determinant in Eq.~(\ref{sigmank1}) of Lemma 3 can be expanded similarly to Eq.~(\ref{sigmanLapB}). Using the above asymptotics and their counterparts for $S_j(\textbf{\emph{x}}^{-}(n,k)+\nu \textbf{\emph{s}}^*)$, and repeating similar calculations as in the proof of Theorem 2, we can readily get
\begin{equation*}
\hspace{-0.8cm}\sigma_{n,k}(x,t) = \hat{\chi}_3 \hspace{0cm} \left|\left[Q_{N}^{[m]}\right]'\hspace{-0.08cm}(z_0)\right|^2 |a_{2m+1}|^{\frac{N(N+1)-2}{(2m+1)}}
\left[ x_1^+(x-\hat{x}_{0}, t-\hat{t}_{0}; n,k ) x_1^-(x-\hat{x}_{0}, t-\hat{t}_{0}; n,k ) +\frac{|p_{1}|^2 }{(p_{0}+p_{0}^*)^2} \right]\left[1+O\left(a_{2m+1}^{-1/(2m+1)}\right)\right],
\end{equation*}
where
\begin{equation*}
\hat{\chi}_3=c_N^{-2} \hspace{0.05cm} \left(\frac{|p_{1}|^2}{(p_{0}+p_{0}^*)^2}\right)^{\frac{N(N-1)}{2}} \left(\frac{2m+1}{2^{2m}}\right)^{\frac{N(N+1)-2}{2m+1}}.
\end{equation*}
From Lemma 3, we see that the leading order term of this asymptotics of $\sigma_{n,k}(x,t)$ yields the fundamental rogue wave of the Manakov system, whose explicit expressions are given in Eqs.~(\ref{u11Mana})-(\ref{u21Mana}), and the error of this leading order approximation is $O\left(a_{2m+1}^{-1/(2m+1)}\right)$. Asymptotics of (\ref{u1NTh3})-(\ref{u2NTh3}) away from the origin in Theorem 3 is then proved.

Regarding the asymptotics (\ref{u1NTh3b})-(\ref{u2NTh3b}) in the neighborhood of the origin, its proof is identical to that for the Boussinesq equation. In particular, the new parameters $\{\hat{a}_1, \hat{a}_3, \cdots, \hat{a}_{2N_0-1}\}$ in the lower-order rogue wave $[u_{1,N_0}(x,t), u_{2,N_0}(x,t)]$ are related to the original parameters $\{a_1, a_3, \cdots, a_{2N_0-1}\}$ by the same relation (\ref{ParameterRela}), which is Eq.~(\ref{akMana}) in Theorem 3. This completes the proof of Theorem 3.

\section{Universal rogue patterns associated with the Yablonskii-Vorob'ev polynomial hierarchy} \label{sec:universal}

Rogue patterns presented in Theorems 1-3 for the GDNLS, Boussinesq and Manakov equations have a lot in common. In all three cases, their rogue waves form clear geometric shapes made up of fundamental rogue waves, whose $(x,t)$ locations are determined by the root structures of the Yablonskii-Vorob'ev polynomial hierarchy $Q_{N}^{[m]}(z)$ through a linear transformation (\ref{x0t0B}). In addition, a possible lower-order rogue wave may appear at the wave center. Since this transformation (\ref{x0t0B}), from the complex root plane of $Q_{N}^{[m]}(z)$ to the $(x,t)$ plane of fundamental-rogue-wave locations, is linear, we see that rogue patterns in these equations are nothing but a linear transformation of the Yablonskii-Vorob'ev root structures. These root structures have the elegant shapes such as triangles, pentagons and heptagons (see Fig.~\ref{f:roots}), and their linear transformations are just those shapes under simple actions such as rotation, dilation, stretch, shear and translation (see Fig.~\ref{images}). This explains all the rogue-wave shapes illustrated in Sec.~\ref{sec:comparison} for these three integrable systems.

Clearly, the above rogue patterns are not restricted to these three integrable systems. In fact, we reported these same rogue patterns first for the NLS equation in Ref.~\cite{YangNLS2021}. In that case, rogue patterns from the fundamental (Peregrine) waves are described by the linear transformation (\ref{x0t0B}) as well, where
\[
\textbf{B}=\left[\begin{array}{cc} \Re(\Omega) & -\Im(\Omega) \\ \Im(\Omega) & \Re(\Omega) \end{array}\right], \qquad \left[\begin{array}{c} \Delta_{1} \\ \Delta_{2} \end{array}\right]=\left[\begin{array}{c} 0 \\ 0 \end{array}\right],
\]
and $\Omega$ is as defined in Eq.~(\ref{defOmega}). We have also checked the three-wave resonant interaction system, whose bilinear rogue waves were derived in Ref.~\cite{YangYang2020-3waves}. Again, for a broad class of its rogue waves which are counterparts of Manakov rogue waves in Lemma 3, we have found that when one of their internal parameters is large, those rogue waves would exhibit geometric shapes, made up of fundamental-rogue-wave ``molecules", that are also the linearly-transformed root structures of the Yablonskii-Vorob'ev polynomial hierarchy, together with a possible lower-order rogue wave at the center, just like the NLS equation and the three other integrable systems considered in this article.

In all these integrable systems, the $\tau$ functions of their rogue waves can be expressed through Schur polynomials with index jumps of 2, see Eqs.~(\ref{sigma_nD}), (\ref{SigmanAlg}), (\ref{sigmank1}) and Ref.~\cite{YangNLS2021,YangYang2020-3waves}. Whenever that happens, we can clearly see from the asymptotic analysis such as that given in the previous section that rogue patterns will always be linearly-transformed Yablonskii-Vorob'ev root structures. Since the $\tau$ functions of rogue waves can also be expressed through Schur polynomials with index jumps of 2 in many other integrable systems, such as the Ablowitz-Ladik equation \cite{OhtaYangAL}, the Yajima-Oikawa system \cite{JCChen2018LS}, the nonlinear Schr\"{o}dinger-Boussinesq equation \cite{XiaoeYong2018}, and the long-wave-short-wave interaction system \cite{Feng2019}, rogue patterns similar to the ones reported in this article would obviously hold for such systems as well. Thus, these are universal rogue patterns associated with the Yablonskii-Vorob'ev polynomial hierarchy.

What are the common features of these universal rogue patterns obtained from a linear transformation (\ref{x0t0B}) to the Yablonskii-Vorob'ev root structures? First, these rogue waves all have geometric shapes such as triangle, pentagon and heptagon, formed by fundamental-rogue-wave components, together with a possible lower-order rogue wave at the center. These geometric shapes are inherited from similar shapes in the Yablonskii-Vorob'ev root structures. Second, the size of the rogue pattern in the $(x,t)$ plane is proportional to $|\Omega|$, i.e., to $|a_{2m+1}|^{1/(2m+1)}$, where $a_{2m+1}$ is the large internal parameter of the rogue wave. The reason is that this size is controlled by the magnitude of the matrix $\textbf{B}$'s elements in the transformation (\ref{x0t0B}), and it is easy to see that the magnitude of every element in the matrix $\textbf{B}$ is proportional to $|\Omega|$. Thirdly, the orientation of the rogue pattern is controlled by the phase of the large parameter $a_{2m+1}$ through the matrix $\textbf{B}$ of the transformation. This third feature has been elaborated for the NLS equation in Ref.~\cite{YangNLS2021}, and is also clearly visible from the matrix $\textbf{B}$'s expressions (\ref{BGDNLS})-(\ref{BMana}) for the GDNLS, Boussinesq and Manakov equations.

Are there differences between rogue patterns in different integrable systems? Definitely, as inspections of rogue patterns displayed in Sec.~\ref{sec:comparison} clearly show. The largest difference is the amount of stretch and shear induced by the matrix $\textbf{B}$ to the Yablonskii-Vorob'ev root structures. For the NLS equation, the Boussinesq equation, the GDNLS equations with $\alpha=1$, and the Manakov system with $p_0=1/2$, the matrix $\textbf{B}$ is equal to a certain scalar multiplying a rotation matrix. In such cases, the action of the $\textbf{B}$ matrix on the root structure is just a combination of dilation and rotation --- no stretch and shear. The resulting rogue patterns are then the most recognizable from the Yablonskii-Vorob'ev root structures, see Figs.~\ref{f:Boussi1}-\ref{f:Boussi2} for the Boussinesq equation and Ref.~\cite{YangNLS2021} for the NLS equation. However, for other integrable systems such as the GDNLS equations with $\alpha\ne 1$ and the Manakov system with $p_0\ne 1/2$, the $\textbf{B}$ matrix does induce stretch and shear, and the resulting rogue patterns are thus skewed versions of the Yablonskii-Vorob'ev root structures, see Figs.~\ref{f:GDNLS1}-\ref{f:GDNLS2} and \ref{f:Mana1}-\ref{f:Mana2}. The degree of this skewness is determined by details of the $\textbf{B}$ matrix. Our studies show that in the GDNLS and Manakov equations, this skewness is often quite visible; but in the three-wave interaction model, this skewness is often more severe.

Another difference between rogue patterns in different integrable systems is the amount of position shifts to the rogue structure, which is induced by the vector $[\Delta_1, \Delta_2]$ in the transformation (\ref{x0t0B}). This position shift is zero for the NLS and GDNLS equations, but nonzero for the Boussinesq and Manakov systems. However, since this position shift is uniform, it does not cause visual changes to the shape of the rogue pattern.

\section{Summary and discussion} \label{sec:summary}

We have shown that universal rogue wave patterns exist in integrable systems. These rogue patterns are linearly transformed Yablonskii-Vorob'ev root structures, with a possible lower-order rogue wave at the center. Visually, these rogue patterns appear in shapes such as triangle, pentagon and heptagon, or their skewed variations due to stretch and shear. They reliably arise when one of the internal parameters in bilinear expressions of rogue waves gets large, and different internal parameters induce different rogue shapes. Detailed analytical predictions of these rogue patterns have been worked out for the GDNLS, Boussinesq and Manakov equations, and these predictions have been found to be in complete agreement with true rogue patterns. When compared to rogue patterns reported earlier for the NLS equation in Ref.~\cite{YangNLS2021}, the main difference is that in these more general integrable systems, rogue patterns can be skewed versions of the Yablonskii-Vorob'ev root structures.

It is noted that these rogue patterns as reported in this article are universal when the $\tau$ functions of the underlying rogue wave expressions can be expressed through Schur polynomials with index jumps of 2, which is the case for the three integrable systems of this article and many others \cite{YangNLS2021,OhtaYangAL,JCChen2018LS,XiaoeYong2018,Feng2019}. However, there also exist rogue waves whose $\tau$ functions cannot be expressed through Schur polynomials with index jumps of 2. For example,  certain rogue waves in the three-wave resonant interaction system as derived by the bilinear method in Ref.~\cite{YangYang2020-3waves} are expressed through Schur polynomials with index jumps of 3 instead of 2. When internal parameters in such rogue waves get large, very different rogue patterns will arise, and they will be asymptotically described by root structures of different types of polynomials. Pattern analysis of such rogue waves is beyond the scope of this article and will be pursued in future publications.

\section*{Acknowledgment}
This material is based upon work supported by the National Science Foundation under award number DMS-1910282, and the Air Force Office of Scientific Research under award number FA9550-18-1-0098.

\begin{center}
\textbf{Appendix A}
\end{center}
In this appendix, we briefly derive the new bilinear expressions of Boussinesq rogue waves presented in Lemma 2. These rogue wave expressions can be obtained by applying the new parameterization developed in Ref.~\cite{YangDNLS2019} to the bilinear derivation of rogue waves in Ref.~\cite{BoussiRWs2020}.
Specifically, instead of the previous choice (7) for the matrix element $m_{ij}$ in Ref.~\cite{BoussiRWs2020}, which we denote as $\phi_{ij}$ in the present paper, we now choose
\[ \label{mijdiff}
\phi_{ij}=\left. \frac{1}{i!}[f(p)\partial_p]^i \frac{1}{j!}[f(q)\partial_q]^j \phi \right|_{p=q=-1},
\]
where
\[ \label{phiBoussi}
\phi=\frac{(p+1)(q+1)}{2(p+q)} \exp\left(\xi+\eta+\sum_{r=1}^\infty a_r \hspace{0.05cm} [\ln \mathcal{W}(p)]^r +\sum_{r=1}^\infty b_r \hspace{0.05cm} [\ln \mathcal{W}(q)]^r \right),
\]
\[
\xi=px_1+p^2x_2, \quad \eta=qx_1-q^2x_2,
\]
$f(p)$, $\mathcal{W}(p)$ are certain well-defined functions given in \cite{BoussiRWs2020}, and $a_r, b_r$ are arbitrary complex constants. Obviously, the function $\tau=\det_{1\le i,j\le N}\left(\phi_{2i-1,2j-1}\right)$ with the above choice of $\phi_{ij}$ also satisfies the bilinear equation (38) in Ref.~\cite{BoussiRWs2020}. Then, when we set $b_r=(-1)^r a_r^*$, $x_1=x/2$ and $x_2=-\textrm{i}t/4$, this $\tau$ function would satisfy the bilinear equation (35) of the Boussinesq equation in Ref.~\cite{BoussiRWs2020}.

Applying the same reduction technique of \cite{BoussiRWs2020} to the above new $\tau$ solution, we can remove the differential operators in the expression (\ref{mijdiff}) of its matrix element $\phi_{ij}$ and reduce it to $\sigma=\det_{1\le i,j\le N}\left(\phi_{2i-1,2j-1}\right)$, where
\begin{equation}
\phi_{i,j} =\sum_{\nu=0}^{\min(i,j)} \left(\frac{-1}{12}\right)^{\nu} \hspace{0.06cm} S_{i-\nu}(\hat{\textbf{\emph{x}}}^{+} +\nu \textbf{\emph{s}})  \hspace{0.06cm} S_{j-\nu}(\hat{\textbf{\emph{x}}}^{-} + \nu \textbf{\emph{s}}),
\end{equation}
vectors $\hat{\textbf{\emph{x}}}^{\pm}=\left( \hat{x}_{1}^{\pm}, \hat{x}_{2}^{\pm},\cdots \right)$ are defined by
\begin{eqnarray}
&& \hat{x}_{r}^{+}=\frac{e^{2\textrm{i}\pi/3}+(-1)^r e^{-2\textrm{i}\pi/3}}{2\cdot 3^{r}\cdot r!} \hspace{0.03cm} \left[x+ (-2)^{r-1} \textrm{i}t \right]+a_{r}, \hspace{1.1cm} r\ge 1,  \\
&& \hat{x}_{r}^{-}=\frac{e^{2\textrm{i}\pi/3}+(-1)^r e^{-2\textrm{i}\pi/3}}{2\cdot 3^{r}\cdot r!} \hspace{0.03cm} \left[ x - (-2)^{r-1} \textrm{i}t \right]+(-1)^r a_{r}^*, \quad r\ge 1,
\end{eqnarray}
and $\textbf{\emph{s}}=(s_1, s_2, \cdots)$ are coefficients from the expansion (\ref{skrkexpcoeffBoussi}) in Lemma 2. Notice that $\left(\hat{x}_{r}^{+}\right)^*=(-1)^r\hat{x}_{r}^{-}$ and $[S_j(\hat{\textbf{\emph{x}}}^{+} +\nu \textbf{\emph{s}})]^*=(-1)^j S_j(\hat{\textbf{\emph{x}}}^{-} +\nu \textbf{\emph{s}})$, and thus $\sigma$ is real. Through a shift of the $x$ and $t$ axes, we normalize $a_1=0$ without loss of generality. Finally, we split the vectors $\hat{\textbf{\emph{x}}}^{\pm}$ into $\textbf{\emph{x}}^{\pm}+\textbf{\emph{w}}^{\pm}$, where $\textbf{\emph{x}}^{\pm}=(\hat{x}_{1}^{\pm}, 0, \hat{x}_{3}^{\pm}, 0, \cdots)$, and $\textbf{\emph{w}}^{\pm}=(0, \hat{x}_{2}^{\pm}, 0, \hat{x}_{4}^{\pm}, \cdots)$. Repeating the same steps as in Appendix A of Ref.~\cite{YangNLS2021}, we can show that the determinant $\det_{1\le i,j\le N}\left(\phi_{2i-1,2j-1}\right)$ becomes one whose matrix element $\phi_{ij}$ is as given in Lemma~2.

\begin{center}
\textbf{Appendix B}
\end{center}

In this appendix, we briefly derive the Manakov bilinear rogue waves presented in Lemma 3.

Under the transformation
\[\label{BilinearTrans}
u_{1}(x,t)=\rho_{1}\frac{g}{f} e^{{\rm{i}} (k_{1}x + \omega_{1} t)},\  \ \ u_{2}(x,t)=\rho_{2}\frac{h}{f} e^{{\rm{i}} (k_{2}x + \omega_{2} t)},
\]
where $f$ is real and $(g, h)$ complex, the Manakov system (\ref{ManakModel1})-(\ref{ManakModel2}) can be converted into the following bilinear equations,
\begin{equation}
\begin{array}{ll}
\left(D_{x}^2 +  \epsilon_{1} \rho_{1}^2 + \epsilon_{1} \rho_{1}^2 \right)f \cdot f =
\epsilon_{1} \rho_{1}^2 g g^*+ \epsilon_{2} \rho_{2}^2 h h^*,\\
\left(i D_{t}+ D_{x}^2 +2i k_{1} D_{x}\right)g \cdot f =0,  \\
\left(i D_{t}+ D_{x}^2 +2i k_{2} D_{x}\right)h \cdot f =0.
\end{array}\label{BiManakov}
\end{equation}
This bilinear system can be reduced from the following higher-dimensional bilinear system in the 2-component Kadomtsev-Petviashvili (KP) hierarchy \cite{OhtaYangWang2011},
\begin{equation}
\begin{array}{ll}
(\frac{1}{2}D_xD_r-1)\tau_{n,k}\cdot\tau_{n,k}=-\tau_{n+1,k}\hspace{0.05cm}\tau_{n-1,k},\\
(D_x^2-D_y+2aD_x)\tau_{n+1,k}\cdot\tau_{n,k}=0,\\
(\frac{1}{2}D_xD_s-1)\tau_{n,k}\cdot\tau_{n,k}=-\tau_{n,k+1}\hspace{0.05cm}\tau_{n,k-1},\\
(D_x^2-D_y+2bD_x)\tau_{n,k+1}\cdot\tau_{n,k}=0,
\end{array} \label{HDManakov}
\end{equation}
where $n, k$ are integers, $\tau_{n,k}$ is a function of four independent variables $(x,y,r,s)$, and
\[ \label{gamma123}
a={\rm{i}}k_{1}, \quad b={\rm{i}}k_{2}.
\]
The solution $\tau_{n,k}$ to these higher-dimensional bilinear equations was given by a certain Gram determinant in Ref.~\cite{OhtaYangWang2011}, but that Gram solution was appropriate only for the derivation of dark solitons. For the derivation of rogue waves here, the solution $\tau_{n,k}$ should be chosen as
\begin{equation} \label{tauMana}
\tau_{n,k}=\det_{1\le \nu, \mu \le N}\Big(\phi_{i_\nu, j_\mu}^{(n,k)}\Big),
\end{equation}
where $(i_1, i_2, \cdots, i_N)$ and $(j_1, j_2, \cdots, j_N)$ are arbitrary sequences of indices,
the matrix element $\phi_{ij}^{(n,k)}$ is defined as
\[
\phi_{ij}^{(n,k)}=\mathcal{A}_i \mathcal{B}_{j}\phi^{(n,k)},
\]
\[
\phi^{(n,k)}=\frac{(p+1)(q+1)}{2(p+q)}\left(-\frac{p-a}{q+a}\right)^{n} \left(-\frac{p-b}{q+b}\right)^{k} e^{\xi+\eta},   \label{phinkMana}
\]
\[
\xi=px+p^2y+\frac{1}{p-a}r+\frac{1}{p-b}s+\xi_{0}(p),
\]
\[
\eta=qx-q^2y+\frac{1}{q+a}r+\frac{1}{q+b}s+\eta_{0}(q),
\]
\[\label{New003b}
\mathcal{A}_{i}=\frac{1}{ i !}\left[f_{1}(p)\partial_{p}\right]^{i}, \quad
\mathcal{B}_{j}=\frac{1}{ j !}\left[f_{2}(q)\partial_{q}\right]^{j},
\]
$p, q$ are arbitrary complex constants, and $\xi_{0}(p), \eta_{0}(q), f_1(p), f_2(q)$ are arbitrary functions of $p$ and $q$ respectively. The reason these functions also satisfy the higher-dimensional bilinear system (\ref{HDManakov}) is that these functions satisfy the same differential and difference relations (10) of Ref.~\cite{OhtaYangWang2011}. Compared to the counterpart choices of the $\tau$-function's matrix elements in Eq.~(8) of Ref.~\cite{OhtaYangWang2011}, our choice of the $\phi^{(n,k)}$ function above contains an extra factor of $(p+1)(q+1)/2$. The benefit of this extra factor is that it will slightly simplify the derivation of rogue wave expressions through Schur polynomials at the end of this appendix. It is noted that this extra factor has also been introduced in the derivation of rogue waves for the Boussinesq equation in Appendix A of this article, and for the NLS equation in Ref.~\cite{YangNLS2021}, for similar reasons.

To reduce the higher-dimensional bilinear system (\ref{HDManakov}) to the original system (\ref{BiManakov}), we need to set
\[
f=\tau_{0,0}, \quad g=\tau_{1,0}, \quad h=\tau_{0,1}, \quad y={{\rm i}}t,
\]
impose the dimension reduction condition
\[ \label{DimMana}
\left(2\partial_x +\epsilon_1\rho_1^2\partial_r+\epsilon_2\rho_2^2\partial_s\right)\tau_{n,k}=C\hspace{0.05cm} \tau_{n,k},
\]
where $C$ is some constant, and impose the conjugation condition
\[ \label{ConjugateMana}
\tau_{-n,-k}=\tau^*_{n,k}.
\]
These two reductions proceed exactly the same way as in Ref.~\cite{YangYang2020-3waves}. Here,
\[
\left(2\partial_x +\epsilon_1\rho_1^2\partial_r+\epsilon_2\rho_2^2\partial_s\right)\phi_{ij}^{(n,k)}=
\mathcal{A}_i \mathcal{B}_{j} \left[\mathcal{F}_{1}(p)+\mathcal{F}_{2}(q)\right]\phi^{(n,k)},
\]
where
\[\label{Q1polynomial}
\mathcal{F}_{1}(p)= \frac{\epsilon_{1}\rho_{1}^2}{p-a} + \frac{\epsilon_{2}\rho_{2}^2}{p-b}+ 2p,
\]
which is the same as the expression (\ref{ManakQ1poly}) in Lemma 3 in view of Eq.~(\ref{gamma123}), and $\mathcal{F}_{2}(q)$ is the above $\mathcal{F}_{1}(p)$ function with $p$ switching to $q$ and $(a,b)$ switching to their complex conjugates. Then, if $p_0$ is a non-imaginary simple root of the algebraic equation
$\mathcal{F}'_{1}(p)= 0$, the dimension reduction and complex conjugation conditions (\ref{DimMana})-(\ref{ConjugateMana}) would be satisfied if we select the $\tau_{n,k}$ determinant (\ref{tauMana}) as
\[ \label{sigmank12}
\tau_{n,k}=
\det_{
\begin{subarray}{l}
1\leq i, j \leq N
\end{subarray} }
\left(
\begin{array}{c}
\phi_{2i-1,2j-1}^{(n,k)}
\end{array}
\right),
\]
where the matrix elements in $\tau_{n,k}$ are defined by
\[ \label{mij-diff}
\phi_{i,j}^{(n,k)}=\left. \mathcal{A}_i \mathcal{B}_{j}\phi^{(n,k)} \right|_{p=p_{0}, \ q=p_{0}^*},
\]
the function $f_1(p)$ inside the operator $\mathcal{A}_i$ is given by
\[ \label{diffoperaf1p}
f_{1}(p)=\frac{\sqrt{\mathcal{F}_{1}^2(p)-\mathcal{F}_{1}^2(p_0)}}{\mathcal{F}'_{1}(p)},
\]
the function $f_2(q)$ inside the operator $\mathcal{B}_i$ is the same as (\ref{diffoperaf1p}) except that the variable subscript 1 changes to 2 and $(p, p_0)$ change to $(q, p_0^*)$, and the function $\eta_{0}(q)$ in $\phi^{(n,k)}$ is taken as $\eta_{0}(q)=[\xi_{0}(p)]^*$. Regarding free parameters, they are introduced through $\xi_{0}(p)$ as \cite{YangDNLS2019,YangYang2020-3waves}
\begin{eqnarray} \label{Thetaxt1}
\xi_0(p)= \sum _{r=1}^\infty  a_{r} \ln^r \mathcal{W}_{1}(p),
\end{eqnarray}
where
\[ \label{W1p}
\mathcal{W}_{1}(p)=\frac{\mathcal{F}_{1}(p)+
\sqrt{\mathcal{F}_{1}^2(p)-\mathcal{F}_{1}^2(p_0)}}{\mathcal{F}_{1}(p_0)},
\]
and $a_{r} \hspace{0.05cm} (r=1, 2, \dots)$ are free complex constants. Notice that $\mathcal{W}_{1}(p)$ is related to $f_1(p)$ through the relation $f_1(p)=\mathcal{W}_{1}(p)/\mathcal{W}'_{1}(p)$, which was the key idea of the $\mathcal{W}$-$p$ treatment for dimension reduction as proposed in Ref.~\cite{BoussiRWs2020} and generalized in Ref.~\cite{YangYang2020-3waves}.

Lastly, we remove the differential operators in the matrix elements (\ref{mij-diff}) and derive more explicit expressions of rogue waves through Schur polynomials. This derivation is very similar to that we did in Ref.~\cite{YangYang2020-3waves} for the three-wave system. In fact, this derivation is a bit simpler now due to our introduction of the extra factor $(p+1)(q+1)/2$ in Eq.~(\ref{phinkMana}). Our algebraic expressions for matrix elements of rogue waves can be further simplified beyond those in Ref.~\cite{YangYang2020-3waves} by using the technique of splitting $\hat{\textbf{\emph{x}}}^{\pm}$ into $\textbf{\emph{x}}^{\pm}+\textbf{\emph{w}}^{\pm}$, just as what we did at the end of Appendix A for the Boussinesq equation. Combining these steps, we then can obtain the rogue wave expressions given in Lemma 3 for the Manakov system. In addition, the parameter $a_1$ can be normalized to zero through a shift of the $(x,t)$ axes.

\end{document}